\documentclass[journal]{IEEEtran}
\usepackage{cite}
\usepackage[inline]{enumitem}
\usepackage{amsmath,amsfonts}
\usepackage{stix}
\usepackage{algorithmic}
\usepackage{graphicx}
\usepackage{array}
\usepackage{textcomp}
\usepackage{stfloats}
\usepackage{url}
\usepackage{verbatim}
\usepackage{textcomp}
\usepackage[hidelinks]{hyperref}
\usepackage[noabbrev, capitalize]{cleveref}
\usepackage[detect-all]{siunitx}
\usepackage{tabularx}
\usepackage{booktabs}
\usepackage{layouts}
\usepackage{subcaption}
\usepackage[numbers]{natbib}
\usepackage[nolist,nohyperlinks]{acronym}

\usepackage{caption}
\usepackage{subcaption}

\usepackage{svg}

\usepackage{tikz}
\usetikzlibrary{calc}
\usetikzlibrary{decorations.pathreplacing,calligraphy}
\usetikzlibrary{shapes.callouts}
\usetikzlibrary{patterns,angles,quotes}
\usetikzlibrary{dsp,chains}
\usepackage{tikzsymbols}
\definecolor{seemoo_organge}{RGB}{206, 101, 0}
\usepackage{pgf-umlsd}

\def\BibTeX{{\rm B\kern-.05em{\sc i\kern-.025em b}\kern-.08em
    T\kern-.1667em\lower.7ex\hbox{E}\kern-.125emX}}
\usepackage{balance}

\renewcommand{\mess}[4][0]{
  \stepcounter{seqlevel}
  \path
  (#2)+(0,-\theseqlevel*\unitfactor-0.7*\unitfactor) node (mess from) {};
  \addtocounter{seqlevel}{#1}
  \path
  (#4)+(0,-\theseqlevel*\unitfactor-0.7*\unitfactor) node (mess to) {};
  \draw[->,>=angle 60] (mess from) -- (mess to) node[midway, above]
  {#3};    
}
\RequirePackage{xargs}
\renewcommandx{\newthread}[4][1=gray!30, 2=0.2]{
  \newinst[#2]{#3}{#4}
  \stepcounter{threadnum}
  \node[below of=inst\theinstnum,node distance=0.8cm] (thread\thethreadnum) {};
  \tikzstyle{threadcolor\thethreadnum}=[fill=#1]
  \tikzstyle{instcolor#3}=[fill=#1]
}

\newif\ifsubmission{}
\submissionfalse{}

\ifsubmission{}
	\newcommand{\TODO}[1]{}
\else
	\newcommand{\TODO}[1]{{\color{red}TODO: #1}}
\fi

\def\BibTeX{{\rm B\kern-.05em{\sc i\kern-.025em b}\kern-.08em
    T\kern-.1667em\lower.7ex\hbox{E}\kern-.125emX}}

\makeatletter
\newcommand\footnoteref[1]{\protected@xdef\@thefnmark{\ref{#1}}\@footnotemark}
\makeatother

\makeatletter
\newcommand{\linebreakand}{%
  \end{@IEEEauthorhalign}
  \hfill\mbox{}\par
  \mbox{}\hfill\begin{@IEEEauthorhalign}
}
\makeatother

\newcommand \titlestring{Smartphones with UWB: Evaluating the Accuracy and Reliability of UWB Ranging}

\begin{document}

\title{\titlestring{}}

\author{
  Alexander Heinrich, Sören Krollmann, Florentin Putz, Matthias Hollick
  \thanks{
    A. Heinrich*, S. Krollmann*, F. Putz, and M. Hollick are with the Secure Mobile Networking Lab in the Computer Science department at TU Darmstadt, Pankratiusstr. 2, 64289 Darmstadt, Germany (e-mail: aheinrich@seemoo.de, skrollmann@seemoo.de, fputz@seemoo.de, mhollick@seemoo.de). \\
    * Both authors contributed equally to this work.
  }
}


\markboth{}
{Alexander Heinrich, Sören Krollmann \MakeLowercase{\textit{(et al.)}}:
\titlestring{}}

\maketitle

\begin{abstract} 
  More and more consumer devices implement the IEEE \ac{uwb} standard to perform distance measurements for sensitive tasks such as keyless entry and startup of modern cars, to find lost items using coin-sized trackers, and for smart payments. While \ac{uwb} promises the ability to perform time-of-flight centimeter-accurate distance measurements between two devices, the accuracy and reliability of the implementation in up-to-date consumer devices have not been evaluated so far. In this paper, we present the first evaluation of \ac{uwb} smartphones from Apple, Google, and Samsung, focusing on accuracy and reliability in passive keyless entry and smart home automation scenarios. To perform the measurements for our analysis, we build a custom-designed testbed based on a \ac{gwen}, which allows us to create reproducible measurements. All our results, including all measurement data and a manual to reconstruct a \ac{gwen} are published online. We find that the evaluated devices can measure the distance with an error of less than \SI{20}{\centi\meter}, but fail in producing reliable measurements in all scenarios. Finally, we give recommendations on how to handle measurement results when implementing a passive keyless entry system.
\end{abstract}



\section{Introduction}\label{sec:introduction}
\IEEEPARstart{W}{ireless} communication is the upcoming norm of interaction between users' phones and other devices such as cars and IoT devices. Without the need to interact with their phone or any other key, a user will be able to unlock their car when approaching and start it up once inside.
At home, smart devices react to our presence. Smart locks open doors when an authorized person is nearby, and smart thermostats regulate temperature based on the presence of owners and their phones. These are only a few examples of the use cases for wireless communication based on \ac{uwb}, the newest wireless physical layer in smartphones, which allows two devices to perform \ac{tof}-based distance measurements. The first car that supports this feature using \ac{uwb} on iPhones was released in 2022~\cite{bmwagBMWAnnouncesBMW}.
However, while a smart light that does not turn on is merely an annoyance, many of the applications like unlocking doors and cars need to be both reliant and accurate to avoid theft or blocked access for the owners.

Different applications have varying needs when it comes to the properties of distance measurements. In general, we can divide them into active and passive use cases. In an active use case, a person wants to actively measure the distance to a certain point, e.g., to find a lost item. Since users are already interacting with their phones in this scenario, they can try to enhance the distance measurements deliberately by holding the smartphone differently or moving around to get a good signal. This is not the case in passive use cases like \ac{pke} and smart home automation systems, which are the focus of our work. In these systems, the smartphones start interactions automatically and the user should not be required to optimize the measurements, as this would defy the idea of a passive system. Instead, the systems need to rely only on the existing measurements and it is currently unclear how measurements behave when a \ac{uwb} smartphone is positioned sub-optimal.

There are two principal use cases of passive distance measurements, which we consider in this work:

\begin{enumerate}
    \item \textbf{A passive keyless entry system} implemented in a car or a smart lock requires accurate distance measurements with an error of less than \SI{50}{\centi\meter}. If the error is larger, the locks may open unpredictably. This may result in an unauthorized person gaining access. 
    Additionally, when the user is closer than \SI{50}{\centi\meter}, the measurements need to be performed reliably and fail at less than \SI{10}{\percent} of cases to ensure a good user experience.
    
    \item \textbf{A smart thermostat, speaker, or light} can offer the ability to change its state depending on the presence of a person in the room. These systems can rely on measurements with accuracy in the order of meters. However, they require reliable measurements at distances of \SIrange{3}{10}{\meter}, because stopping the music or turning off the heating while the person is still present can be disruptive to the user experience. 
To achieve high reliability, good software support for apps on a smartphone is essential. Smart home apps need to be able to connect to devices and start \ac{uwb} ranging seamlessly and reliably without disturbing the user and without failing.
\end{enumerate}

Manufacturers of \ac{uwb} chipsets advertise their devices as producing errors of less than $\pm \SI{10}{\centi\meter}$~\cite{nxpsemiconductorsSR040UltraWidebandTransceiver2021,qorvoinc.DW3000DataSheet2020}. However, at of the time of writing, no publicly accessible evaluation of smartphones with \ac{uwb} capabilities exists. 
We do not know how accurate and reliable \ac{uwb} in a smartphone is.
This raises a couple of issues that need to be addressed: 
\begin{itemize}
    \item \textbf{Reliability and Accuracy of \ac{uwb} technology.} People using \ac{uwb} as a \ac{pke} system need to know if they can rely on accurate measurements. Without this, people will be less likely to adopt the technology and more likely to resist its implementation. A positive evaluation will provide peace of mind to people using a \ac{uwb}-smartphone as a car key.
    \item \textbf{Implementation of UWB measurements.} Manufacturers of IoT devices need to understand how accurately and reliably measurements are reported by smartphones to implement dependable algorithms.
    \item \textbf{Security.} By design, UWB is secure against relay and signal amplification attacks. However, if paired with highly inaccurate and strongly fluctuating measurements the security gain is minimal. It, therefore, needs to be evaluated how current implementations are behaving. Many \ac{pke} systems of cars without \ac{uwb} have been successfully attacked~\cite{staatAnalogPhysicalLayerRelay2022,ibrahimKeyAirHacking2019} and it is questioned if \ac{uwb} \ac{tof} measurements can solve the existing problems.
\end{itemize}

Our measurements and their evaluation address these concerns by providing data and a reproducible testbed to analyze future devices. The goal of this paper is to evaluate \ac{uwb} in smartphones and to analyze if the measurements meet the requirements defined above of \ac{pke} systems and IoT smart home devices. To achieve this, we design and implement a GWEn testbed that allows the rotation of a \ac{dut} by 360º in two axes. A manual on how to build our GWEn testbed is available online and the required software is open-source~\cite{krollmannGWEnGimbalBased2022}. Researchers can assemble it to reproduce our measurements or evaluate new devices.

Utilizing our testbed, we perform reproducible \ac{uwb} measurements with one smartphone each from Apple, Google, and Samsung and a DW3000 development kit from Qorvo. All measurements are performed in three different environments: Outside to measure a ground truth, in our lab that resembles an office environment used to evaluate smart home automation, and in a parking garage to evaluate \ac{pke} in cars. Each environment represents a real-world scenario in which \ac{uwb} can be used for distance measurements. We present our results and cover the differences between each environment. All our measurement data is available online~\cite{heinrich_alexander_2023_7702153}. In addition, we discuss if \ac{uwb} in smartphones can perform well enough to achieve the defined goals in accuracy and reliability, we compare our results against previous work which focused on \ac{uwb} development kits, and we give recommendations for implementing a secure \ac{pke} system.  


Our evaluation reveals the following:
\begin{enumerate}
    \item \ac{uwb} measurements are not reliable. Our measurements show that \ac{uwb} antennas embedded in smartphones are highly directional. This directionality is caused by the internals of a smartphone (battery, main board, display, etc.) which shield the signal in one direction. 
    \item Measurements are accurate enough for \ac{pke} systems. On average the devices measure an error of less than \SI{21}{\centi\meter}. This is independent of the smartphone’s orientation. 
    \item \ac{uwb} measurements produce outliers of several meters, which influences the accuracy of the measurements. However, the overall standard deviation of our measurements is generally less than \SI{25}{\centi\meter}. 
    \item The software integration of \ac{uwb} into the Android OS is unstable, especially on Google Pixel smartphones. Samsung's devices achieve better overall stability, notably when using their internal \ac{uwb} API.
\end{enumerate}


Our key contributions are: 
\begin{enumerate}
    \item{We are the first to perform measurements with \ac{uwb} capable Apple, Samsung, and Google smartphones.}
    \item{We design and build \ac{gwen}, a reproducible 3D-printed testbed that can rotate a \ac{dut} in two axes around the \ac{uwb} antenna for full \SI{360}{\degree}.}
    \item{We present and evaluate our measurements on accuracy and reliability to show strengths and weaknesses of \ac{uwb} ranging.}
    \item{We give recommendations to manufacturers to implement secure entry systems.}
    \item All our measurement data, that we evaluate in this paper is available online. 
    \item A manual on how to build a \ac{gwen} testbed and the required software are open-source.
\end{enumerate}

\section{Background \& Related Work}\label{sec:background}
\noindent This section gives an introduction to \ac{uwb}, the available hardware, applications, and how 3D printing can be used to allow reproducible measurements.
According to the IEEE standards, \ac{uwb} defines two modes: \ac{hrp} and \ac{lrp}. In our work, we focus on \ac{uwb} \ac{hrp} as it is the only mode supported by smartphones and used for \ac{pkes} systems in cars.  

\subsection{UWB HRP Mode}\label{sec:background_uwb_hrp}
Most systems that are currently deployed use the \ac{uwb} \ac{hrp} mode that has been first defined in 802.15.4a--2007\cite{ieee802154-2007}.
Over the years, this mode has been integrated into the main standard, and it has been extended at last by 802.15.4--2020 and 802.15.4z. While 802.15.4--2020 added new data rates and \acp{prf}, 802.15.4z added additional security mechanisms that should protect ranging measurements against external attacks (see \cref{sec:background_uwb_security}). 
We explain the distinctive features of \ac{uwb}, how to use it for ranging and locating devices, and its security properties. 

\subsubsection{Physical Layer}
\begin{figure}[!t]
    \centering

    \fontsize{4}{4}\selectfont        
    \includegraphics[width=.8\linewidth]{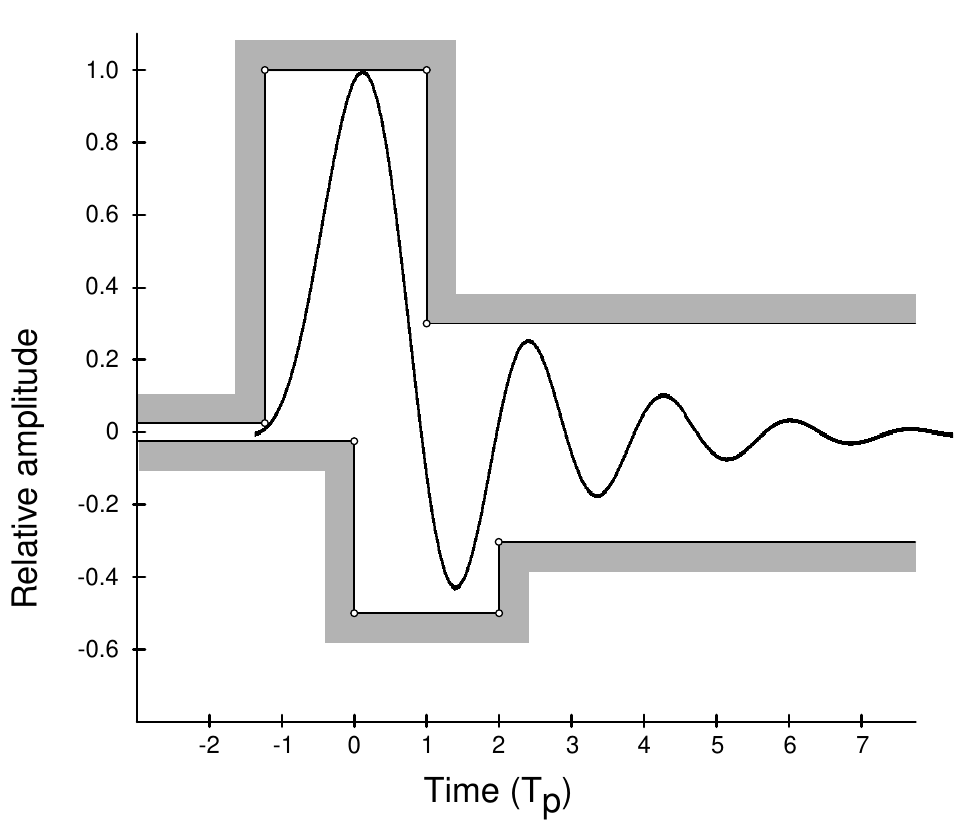}
    
    \caption{Recommended time domain mask for the UWB HRP PHY pulse~\cite{IEEEStandardLowRate2020z}.}\label{fig:uwb_pulse}
\end{figure}

The discussed \ac{uwb} systems are pulse-based communication systems. On the physical layer, these systems send out short electromagnetic pulses instead of longer sine waves. These pulses have sharp edges to the maximum amplitude and a total duration of \SIrange[]{0.74}{2}{\nano\second}~\cite{IEEEStd8022020}.
An example of such a pulse is visible in \cref{fig:uwb_pulse}.
These sharp pulses allow accurate detection of the \ac{toa} of the signal at the receiver.
The U1 chipset in Apple's devices has a clock that allows a theoretical resolution of \SI{1}{\pico\second}\footnote{Retrieved from reverse-engineering iOS.}. Chipsets from Decawave offer up to \SI{15.65}{\pico\second}~\cite[p.32]{qorvoinc.DW3000DEVICEDRIVER2019}. 

\ac{uwb} \ac{hrp} supports 15 different channels with different bandwidths between \SI{499.2}{\mega\hertz} and \SI{1354.92}{\mega\hertz}. Complex channels, formed by prepending a specific preamble to each frame, allow concurrent transmissions on the same channel~\cite{IEEEStd8022020}. 

The \ac{hrp} mode defines three mean \ac{prf} at \SI{3.9}{\mega\hertz}, \SI{15.6}{\mega\hertz} and \SI{62.40}{\mega\hertz}. All devices that we were testing used \SI{62.40}{\mega\hertz}. Higher \acs{prf} achieve higher data rates and faster ranging. 
The \ac{uwb} standard IEEE 802.15.4--2020 also defines an \ac{lrp} mode which uses a lower \ac{prf} and a different security paradigm. We do not cover the details of this mode in this paper. There is currently no smartphone that supports \ac{lrp}.  


\begin{figure}[!t]
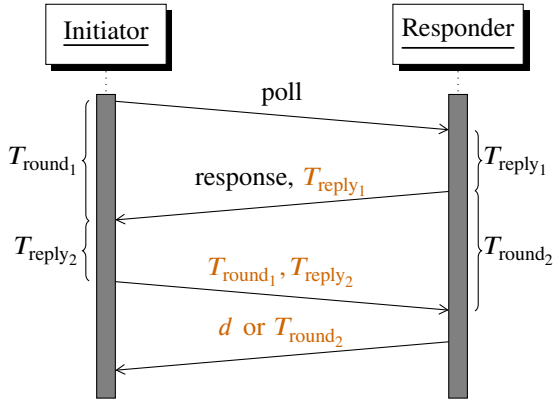

    \centering
    \begin{sequencediagram}
        \renewcommand\unitfactor{0.4}
        \newthread[gray][0]{i}{Initiator}
        \newthread[gray][3]{r}{Responder}
        
        \mess[1]{i}{poll}{r};
        \node[anchor=east] (t0) at (mess from) {};
        \node[anchor=west] (t1) at (mess to) {};

        \postlevel
        \mess[1]{r}{response, \color{seemoo_organge}$T_{\text{reply}_1}$}{i};
        \node[anchor=west] (t2) at (mess from) {};
        \node[anchor=east] (t3) at (mess to) {};

        \postlevel
        \mess[1]{i}{\color{seemoo_organge}$T_{\text{round}_1}, T_{\text{reply}_2}$}{r};
        \node[anchor=east] (t4) at (mess from) {};
        \node[anchor=west] (t5) at (mess to) {};

        \begin{scope}[]
            \mess[1]{r}{\color{seemoo_organge}$d$ or $T_{\text{round}_2}$}{i};
        \end{scope}

        \draw[decorate,decoration = {brace}] ($(t1.center)- (-0.1,0)$) -- ($(t2.center)- (-0.1,0)$)
        node [align=center,midway,anchor=west] {$T_{\text{reply}_1}$};

        \draw[decorate,decoration = {brace}] ($(t3.center)- (0.1,0)$) -- ($(t0.center)- (0.1,0)$)
        node [align=center,midway,anchor=east] {$T_{\text{round}_1}$};

        \draw[decorate,decoration = {brace}] ($(t2.center)- (-0.1,0)$) -- ($(t5.center)- (-0.1,0)$)
        node [align=center,midway,anchor=west] {$T_{\text{round}_2}$};

        \draw[decorate,decoration = {brace}] ($(t4.center)- (0.1,0)$) -- ($(t3.center)- (0.1,0)$)
        node [align=center,midway,anchor=east] {$T_{\text{reply}_2}$};

    \end{sequencediagram}
    \caption{Message exchange diagram of Double Sided - Two Way Ranging. After the third message the responder is able to calculate the distance using \Cref{eq:dist_ds}. Elements in organge are optional.}\label{fig:seq_ds_twr}
\end{figure}
\subsubsection{Ranging mechanisms}\label{sec:background_uwb_ranging}
To perform ranging, the \ac{uwb} \ac{hrp} transceivers use \ac{tof} based distance measurements. 
By measuring the time it took the signal to travel from the transmitter to the receiver, it is possible to calculate the distance between both devices. To remove the necessity of clock-synchronization between both devices, a \ac{twr} mechanism is used. 
\Ac{ds-twr} is used to directly correct clock drifts and to ensure that both devices have a measurement result available.
The \ac{ds-twr} is depicted in \cref{fig:seq_ds_twr} and works as follows:
The initiator starts by sending a poll message to the responder. The responder responds and also transmits the reply delay ($T_{\text{reply}_1}$) to the initiator.
The initiator now sends a second poll message and transmits its reply delay ($T_{\text{reply}_2}$) and overall roundtrip time for the first message exchange ($T_{\text{round}_1}$) to the responder. The responder now calculates the final distance by using \cref{eq:dist_ds}. In a last (optional) message, the responder transmits the measured distance ($d$) or the second roundtrip time ($T_{\text{round}_2}$). 

\begin{equation}
    \label{eq:dist_ds}
    d=\frac{T_{\text{round}_1} * T_{\text{round}_2} - T_{\text{reply}_1} * T_{\text{reply}_2}}{T_{\text{round}_1} + T_{\text{round}_2} + T_{\text{round}_1} + T_{\text{round}_2}}*c
\end{equation}

\subsubsection{Security Properties}\label{sec:background_uwb_security}

Earlier versions of \ac{uwb} \ac{hrp} in 802.15.4 did not have any security properties. The physical layer definitions were only supplemented by a simple MAC layer without any protection of ranging measurements against malicious interference. 

This allowed severe attacks against \ac{uwb} if used for ranging and localization. By listening to the \ac{uwb} channel, an attacker can easily identify the used preamble, which starts every \ac{uwb} frame. The early detection of the preamble allows an attacker to send frames with the same preamble earlier and thus reduce the measured distance~\cite{fluryEffectivenessDistancedecreasingAttacks2010}. Even sending static \ac{uwb} pulses at a fixed interval to interfere with any surrounding ranging measurements caused distance reduction and enlargement attacks~\cite{poturalskiCicadaAttackDegradation2010}. 


The current standard amendment IEEE 802.15.4z introduced a new protection mechanism against physical layer attacks~\cite{IEEEStandardLowRate2020z}. 
Each \ac{uwb} \ac{hrp} frame now contains a \ac{sts}. The \ac{sts} is a cryptographically secured sequence of pulses that is derived from a pseudo-random generator and a shared secret between the initiator and responder. Both devices in a ranging session will only accept frames that contain an expected \ac{sts} and the location of the \ac{sts} in the frame is used to derive accurate and secured timestamps. 

An attacker that tries to reduce the measured distance by sending a packet with the same preamble on the same channel earlier does not succeed, because the \ac{sts} changes with every frame and cannot be known in advance by an attacker. The receiver simply discards the frame. Previous research has shown that a distance reduction is still possible, but only at a low success rate (see \cref{sec:related_work})~\cite{279984}.

\subsection{Available Hardware}

We briefly introduce the available consumer hardware for \ac{uwb} \ac{hrp} devices in this section. There are currently three main manufacturers of chipsets in smartphones: Apple, NPX, and Qorvo. \Cref{tab:available_uwb} gives an overview of available hardware. 

The iPhone 11 in 2019 was the first mass-market smartphone that integrated a \ac{uwb} \ac{hrp} chip to perform ranging~\cite{U1ChipIPhone2019}. The previously released Bespoon phone has only been a public prototype, which is no longer on sale~\cite{stmicroelectronicsn.v.HISTORYUWB2020}. The Apple U1 \ac{uwb} chip has been deployed to a number of devices: iPhone, Apple Watch, AirTag, and HomePod mini. The iPhone is the only device that has a multi-antenna \ac{uwb} chip allowing \ac{aoa} measurements.  

Samsung was the second manufacturer which integrated \ac{uwb} in their high-end Galaxy Note, Plus and Ultra lines~\cite{samsungelectronicsincSamsungUnveilsFive2020}. They integrated the NXP SR100T chips into their Galaxy smartphones. The Samsung SmartTag+ integrates a precision finding feature leveraged by \ac{uwb} and the NXP SR040 chip. 
While the SR100T chip also integrates \ac{aoa} by using multiple antennas, the SR040 chip only uses one antenna and is designed for low-cost systems~\cite{nxpsemiconductorsSR040UltraWidebandTransceiver2021}. 

The Google Pixel 6 Pro was the first to introduce \ac{uwb} to Google's phones. This phone implements a \ac{uwb} chipset from Qorvo. The chipset is based on the DWM3000 family of chips~\cite{qorvoinc.QorvoDeliversUltraWideband}. The Pixel 6 Pro also supports \ac{aoa} measurements. 

All chipsets are compliant with the open IEEE 802.15.4z standard and the proprietary \ac{fira} and  \ac{ccc} standards. This allows theoretically wide interoperability between these chips. However, due to limited software support, Android and iOS smartphones are not able to interoperate, yet~\cite{NearbyInteractionApple}.
In this work, we focus on smartphones with \ac{uwb} support: The Apple iPhone (11 and newer), Samsung Galaxy smartphones (S21 Ultra and newer), and Google Pixel (6 Pro and newer).

\begin{table*}[]
    \begin{minipage}{\textwidth}
        \caption{Commercially available UWB chips evaluated in this work}\label{tab:available_uwb}
        \begin{tabularx}{\linewidth}{llllll}
            \toprule
        Device               & Chipset      & No. Antennas & Supported Standards              & Accuracy\footnote{As stated by chip manufacturer~\cite{mobileknowledgeMKUWBKit,qorvoinc.DW3000DataSheet2020}} & Channel    \\
            \midrule
            
        Apple devices         & Apple U1     & $1-3$            & IEEE 802.15.4z, FiRa\footnote{\label{fira}FiRa Consortium~\cite{FiRaConsortium}}, CCC\footnote{\label{ccc}Car Connectivity Consortium~\cite{CarConnectivityConsortiuma}}, ANI\footnote{\label{ani}Apple Nearby Interaction~\cite{NearbyInteractionApple}}  & -         & 5, 9       \\

        Samsung Galaxy smartphones & NXP SR100T    & 3           & IEEE 802.15.4z, FiRa\footnoteref{fira}, CCC\footnoteref{ccc} &   $\pm$\SI{10}{\centi\metre}       & 5, 6, 8, 9          \\
        
        Google Pixel Pro     & Qorvo DW3720 & 2            & IEEE 802.15.4z, FiRa\footnoteref{fira}, CCC\footnoteref{ccc}        & $\pm$\SI{10}{\centi\metre}  & 5, 6       \\

        Qorvo DWM3000EVB     & DW3110       & 1            & IEEE 802.15.4z, FiRa\footnoteref{fira}, CCC\footnoteref{ccc}, ANI\footnoteref{ani}   & $\pm$\SI{10}{\centi\metre}  & 5, 6       \\

        \bottomrule

        \end{tabularx}
\end{minipage}
\end{table*}

\subsection{Applications}
Smartphones with \ac{uwb} have a few applications in which the technology is used to provide an extra layer of security or positional awareness. We present major applications that are already deployed. 

The Digital Key 3.0 allows hands-free keyless access using distance measurements of \ac{uwb} in combination with Bluetooth Low Energy to check whether an authorized person is near the vehicle~\cite{CarConnectivityConsortium}. The vehicle unlocks automatically and can be started when the person sits inside (\ac{pkes}). At the time of writing only iPhones support \ac{uwb} car keys, while the Google Pixel offers support for \ac{nfc}-based car keys. 

NXP and the ING bank announced that they will work on supporting \ac{uwb} on Samsung smartphones to use it for payments to nearby users. \Ac{uwb} is intended as an extra security layer and omits the need to exchange bank account numbers before sending a payment~\cite{inggroupPointPay2022,nxpsemiconductorsNXPCollaboratesING}.

\Ac{uwb}-tags are already widespread and allow the tracking of personal belongings.
These tags are usually not much bigger than a coin, have an integrated power-saving \ac{uwb} chip and can therefore run for several months on one charge. The precision finding feature uses \ac{uwb} to measure the distance and the \ac{aoa} to the missing tracker and shows that on the user's smartphone~\cite{NearbyInteractionApple,samsungelectronicsincIntroducingNewGalaxy2021,appleinc.FindYourKeys2022}.

Several smart home and \ac{iot} applications would benefit from the application of \ac{uwb}. A smart door lock with \ac{uwb} would increase the security over a \ac{ble}-based version. In November 2022, Samsung announced plans for an aforesaid smart lock~\cite{shaikSamsungWalletGets2022}. 
Presence detection of people in a room could be leveraged to control lighting and heating based on personal preferences. To the best of our knowledge, no smart home devices with \ac{uwb} are available on the consumer market.

\subsection{Related Work}\label{sec:related_work}

Many works focus on indoor-localization techniques based on \ac{uwb}~\cite{niculescuEnergyefficientPreciseUWBbased2022,flueratoruHighAccuracyRangingLocalization2022, caoDistributedTDMAMobile2021, fengKalmanFilterBasedIntegrationIMU2020}.
Our work focuses on consumer smartphones, their \ac{uwb} ranging accuracy, reliability, and the security implications that inaccurate measurements can have. Therefore, the related work can be separated into two areas, the general \ac{uwb} accuracy and the security analyses of the \ac{uwb} protocols. To the best of our knowledge, we are the first to evaluate and compare the \ac{uwb} ranging accuracy of consumer smartphones. 

\subsubsection{Ranging accuracy}

Related work in this area mainly concentrates on the evaluation of available development kits for \ac{uwb} hardware. Between 2015 and 2017, after the first \ac{uwb} chipsets became available, the first evaluations on ranging accuracy and performance have been conducted~\cite{malajnerUWBRangingAccuracy2015,jimenezComparingDecawaveBespoon2016,jimenezruizComparingUbisenseBeSpoon2017}. During that time only the \ac{uwb} \ac{hrp} mode was available, and it was lacking any security measures. All devices evaluated during this time are no longer compatible with current \ac{uwb} standards.
In 2020, after the release of IEEE 802.15.4z with \ac{lrp} and the \ac{sts} to secure \ac{hrp} measurements, studies have started to compare \ac{lrp} and \ac{hrp} in terms of energy consumption~\cite{flueratoruEnergyConsumptionRanging2020}. 
Works in 2022 also evaluate the accuracy of the DW3000 chipsets from Qorvo in development kits, like the DWM3000EVB~\cite{tiemannExperimentalEvaluationIEEE2022a,flueratoruHighAccuracyRangingLocalization2022,juranHandsOnExperienceUWB2022}.
The DW3000 has also been compared to \ac{lrp} devices~\cite{flueratoruHighAccuracyRangingLocalization2022}. In 2022, researchers discovered that the default DW1000 antenna has an uneven radiation pattern and the researchers evaluated the ranging error from several angles by using unmanned aerial vehicles \cite{niculescuEnergyefficientPreciseUWBbased2022}. 
We compared several works with our result in~\cref{ssec:comparison_related_work}. 

\subsubsection{UWB Security}
IEEE 802.15.4z \ac{hrp} mode has been evaluated for its security and the first attacks on it have been discovered. Previous work analyses the theoretical potential for multipath effects that could cause a distance reduction attack~\cite{singhSecurityAnalysisIEEE2021a}. Since no open-source implementations are available, a simulated \ac{hrp} receiver according to the standard has been implemented and evaluated. Follow-up work has confirmed that the theoretical attack also works on devices that integrated Apple's U1 \ac{uwb} chip~\cite{279984}. Practical distance reductions of several meters have been possible by using a DWM3000 \ac{uwb} sending frames with low power and the same preamble at the same time as the other \ac{uwb} transmitter.
The success rate of the attack was less than \SI{5}{\percent}. Hence, it shows that the applied security measures are not sufficient to stop all attacks.


\makeatletter
\AC@reset{gwen}
\makeatother

\section{GWEn}\label{sec:gwen}
\noindent 
The main goal of this work is to determine the ranging accuracy and reliability of \ac{uwb} smartphones in a standardized, repeatable, reproducible, and automated way.
This enables a systematic evaluation and comparison of our \acp{dut}.

\begin{figure}[!t]
    \centering
    \includegraphics[width=\linewidth]{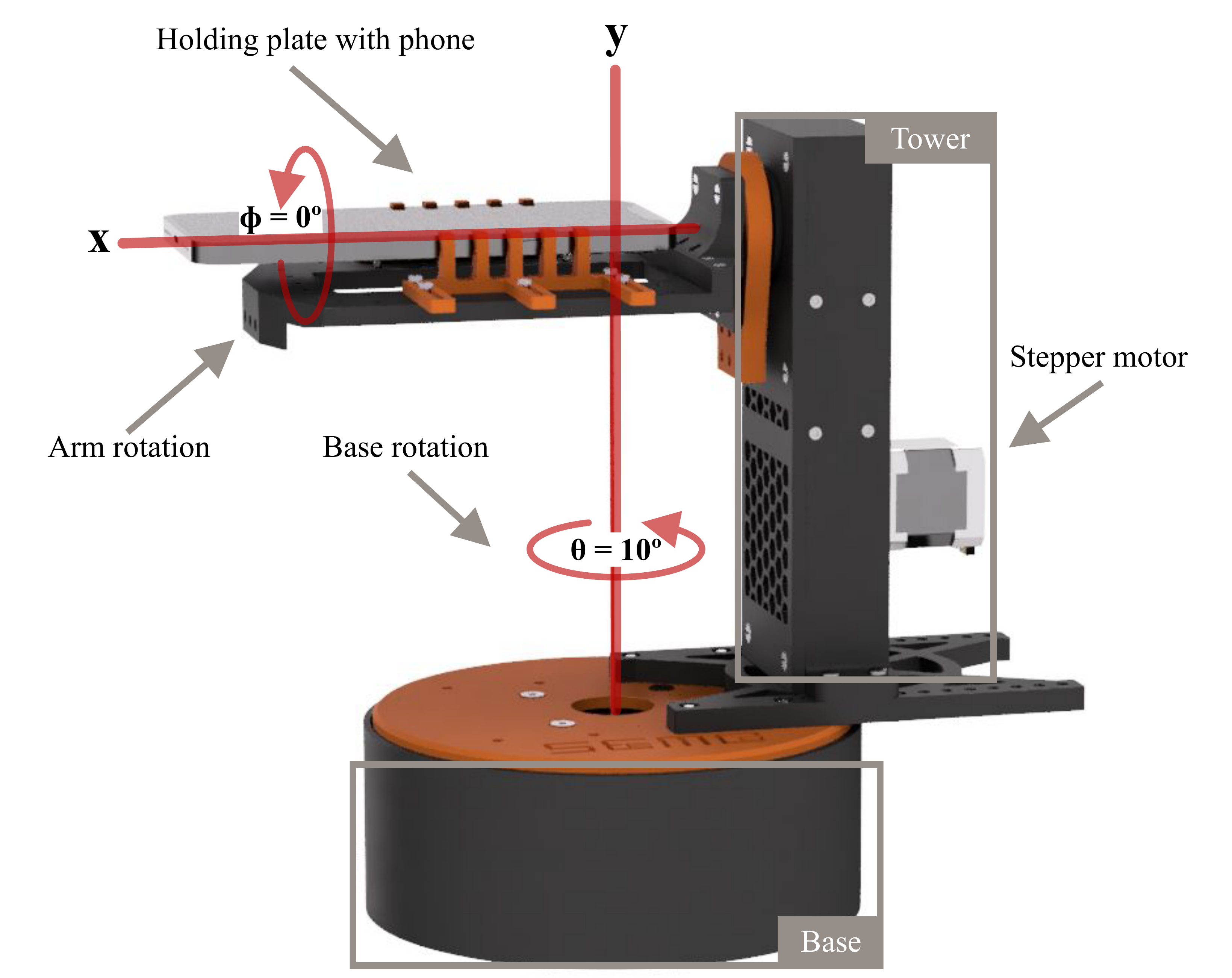}
    \caption{$3$D rendering of an assembled build of GWEn, highlighting the two rotation axes in red. Both axes can rotate independently for a full 360° each.}\label{fig:gwen_rotation_axes}
\end{figure}

To accomplish this, we develop, build, and evaluate a \ac{gwen}. \Cref{fig:gwen_rotation_axes} shows a labeled build of \ac{gwen}. 
It is equipped with two rotation axes, one in the base and one in the arm as depicted in \cref{fig:gwen_rotation_axes}.
Throughout this paper, the angle $\phi$ denotes the arm rotation around the x-axis and the angle $\theta$ denotes the base rotation around the y-axis. 

\subsection{Measurements with GWEn}
\Ac{gwen} is a multipurpose tool for conducting wireless measurements and evaluating antennas at a small scale. 

Each axis can rotate the full 360° making it possible to position the \ac{dut}, attached to the holding plate, in any desired orientation with a resolution of $\sim$0.01°.
A variety of devices can be attached to the holding plate, allowing all kinds of wireless transmission measurements or antenna evaluations. 
For an optimal measurement, the antenna of the \ac{dut} must be as close as possible to the intersection of the two rotation axes.
The tower on the rotation plate can be moved further outward or inward to accommodate devices of different lengths.
The holding plate on the tower can be moved further up or down to adapt \ac{gwen} to devices of different thicknesses.
To be able to extract the measurement data from the \ac{dut}, it is connected to \ac{gwen} via a USB connection.

\ac{gwen} can be controlled over a web interface that is reachable via its own Wi-Fi hotspot. Furthermore, it's also possible to create custom Python scripts that allow different movement patterns. 

To create meaningful and repeatable measurements, we have developed a workflow for \ac{gwen} measurements:
\begin{enumerate}
    \item \ac{gwen} setup: This includes setting up \ac{gwen} at the test location, mounting the \ac{dut} and setting up the measurement parameters.
    \Cref{ssec:general_setup} goes into more detail on the specific setup used in our work.
    \item Start the measurement.
    From now on, \ac{gwen} will automatically move the \ac{dut} to previously specified orientations, take the desired number of measurements, and store them accordingly.
    \item A measurement between two iPhones with the settings described in \cref{ssec:general_setup} takes $\sim$90 minutes. No monitoring of the system is required during this period. 
    \item Download the measurement.
    After finishing the measurements, the recordings can be downloaded as zip files.
    Our evaluation software directly works on the generated zip files and generates graphs to evaluate the performance. 
\end{enumerate}

\subsection{Manufacturing}\label{ssec:gwen_manufacturing}
We designed \ac{gwen} to be $3$D-printed using a fused deposition modeling $3$D printer.
Besides belts, screws, bearings and the electrical components all parts are $3$D-printed.
We use \ac{pla} as a printing material. \Ac{pla} has only very little influence on electromagnetic signals \cite{boussatourDielectricCharacterizationPolylactic2018}.
\SI{1}{\centi\metre} of solid \ac{pla} adds about \SI{0.6}{\centi\metre} of measured distance to an \ac{uwb} signal.
We evaluate the interference introduced by \ac{gwen} with more detail in \cref{sec:eval_gwen}.

$3$D-printing is a low-cost and widely available option that allows easy reproduction of desired parts.
Furthermore, it enables modifications, as new parts can be designed and $3$D-printed rapidly.
It takes about $4.5$ days to print all parts on a Prusa i3 MK3S+~\cite{prusaresearcha.s.OriginalPrusaI3}. 
However, this time depends heavily on the printer and the print settings.
Assembling \ac{gwen} and soldering up the electrical components requires additional $\sim$80 hours of manual work.

The parts that cannot be $3$D-printed are off-the-shelf parts.
In total, the material cost for \ac{gwen} was $\sim$410~€.
We released the $3$D files, a list with all additional components required, detailed build instructions, as well as wiring diagrams and source code in our Zenodo repository~\cite{krollmannGWEnGimbalBased2022}.

The software of \ac{gwen} consists of three individual parts:
\begin{enumerate*}
    \item ~the~\textit{web interface} written in Python,
    \item ~the~\textit{measurement software} written in Python, and
    \item ~the~\textit{hardware controller} written in C/C++.
\end{enumerate*}
This modularity allows anyone to rebuild \ac{gwen} as well as expand and adapt it to individual needs.

\subsection{Measurement recording}\label{ssec:sources}

We programmed \ac{gwen} to interact with a variety of devices (see \cref{tab:available_uwb}) and process their \ac{uwb} measurements.
Since the devices come from different manufacturers and run with different software, \ac{gwen} offers the possibility to add new communication options with the help of \textit{sources}.
Each \textit{source} is a plugin written in Python that has to be defined once and provides functions for \ac{gwen} to extract the necessary data from the device.
For smartphones, we use device logs to extract ranging data from the device (see \cref{ssec:dev_conf}).
The \textit{source} to be used for each measurement can be specified at the start of a measurement.

At the end of each measurement, \ac{gwen} creates a recording file.
It contains a JSON file with the setup settings used to create the recording as well as all made distance measurements ordered by their respective base and arm angle.
It is furthermore possible to add additional files, e.g., a complete log of the measurement.
All measurement data is backed up constantly and can be restored.


\section{Evaluation of GWEn}\label{sec:eval_gwen}
\noindent 
As described in \cref{ssec:gwen_manufacturing}, a 1 cm thick \ac{pla} wall has a negligible effect on the signal.
In addition to the parts printed from \ac{pla}, \ac{gwen} also contains motors, ball bearings and shafts.
These parts consist of metal that interferes with the signal between the two devices.
This is especially the case when the tower is directly between the two antennas.
Therefore, before evaluating the data, it must be ensured that the measurement is not affected by \ac{gwen}'s test setup and procedure.

\subsection{Maximum error}

\begin{figure}[!t]
  \centering
  \begin{subfigure}{.48\linewidth}
    \centering
    \includegraphics[width=.98\linewidth]{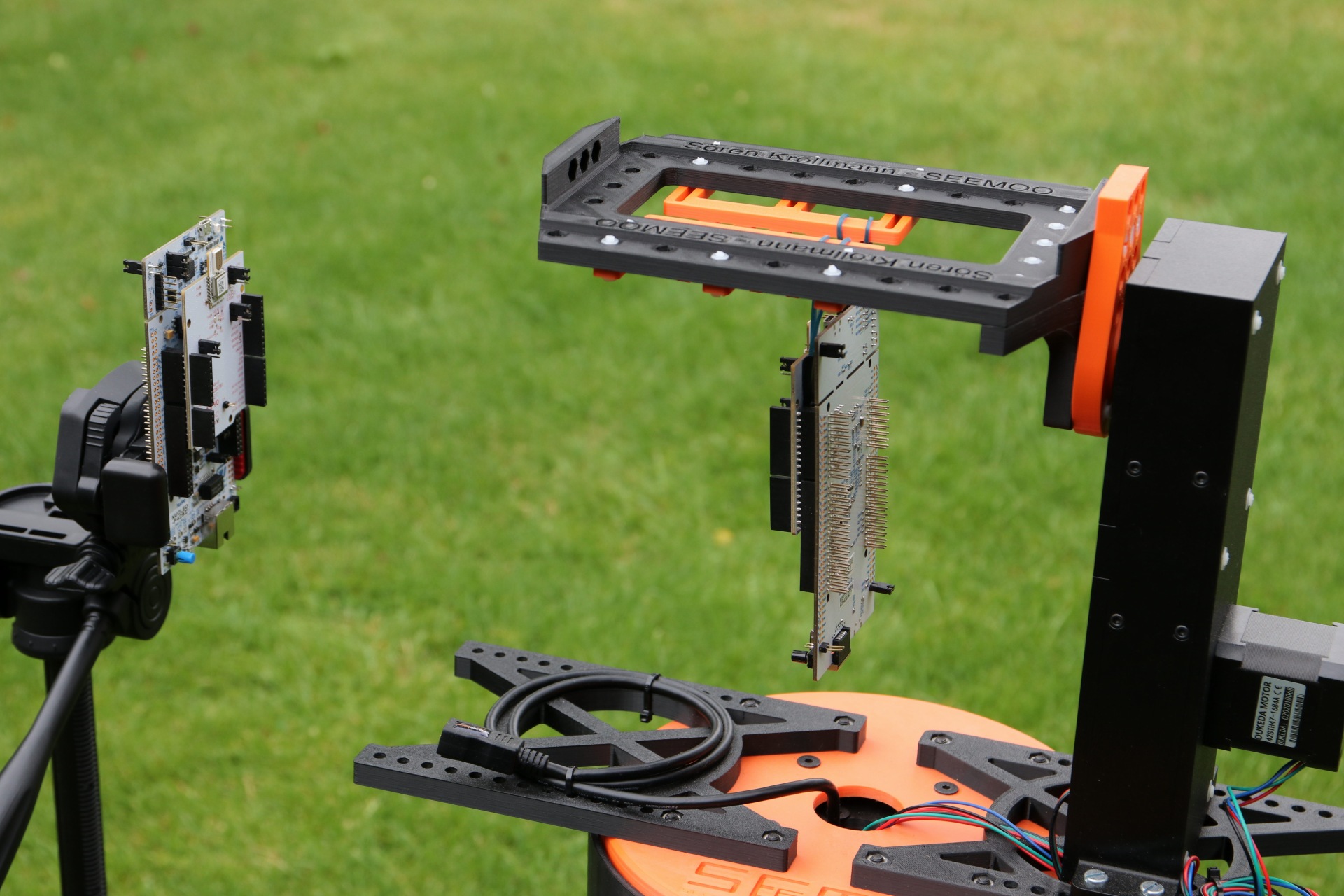}
    \caption{\raggedright{} LOS measurement without obstructions at a base rotation of $\theta=\SI{0}{\degree}$.}\label{fig:t1}
  \end{subfigure}%
  \hfill
  \begin{subfigure}{.48\linewidth}
    \centering
    \includegraphics[width=.98\linewidth]{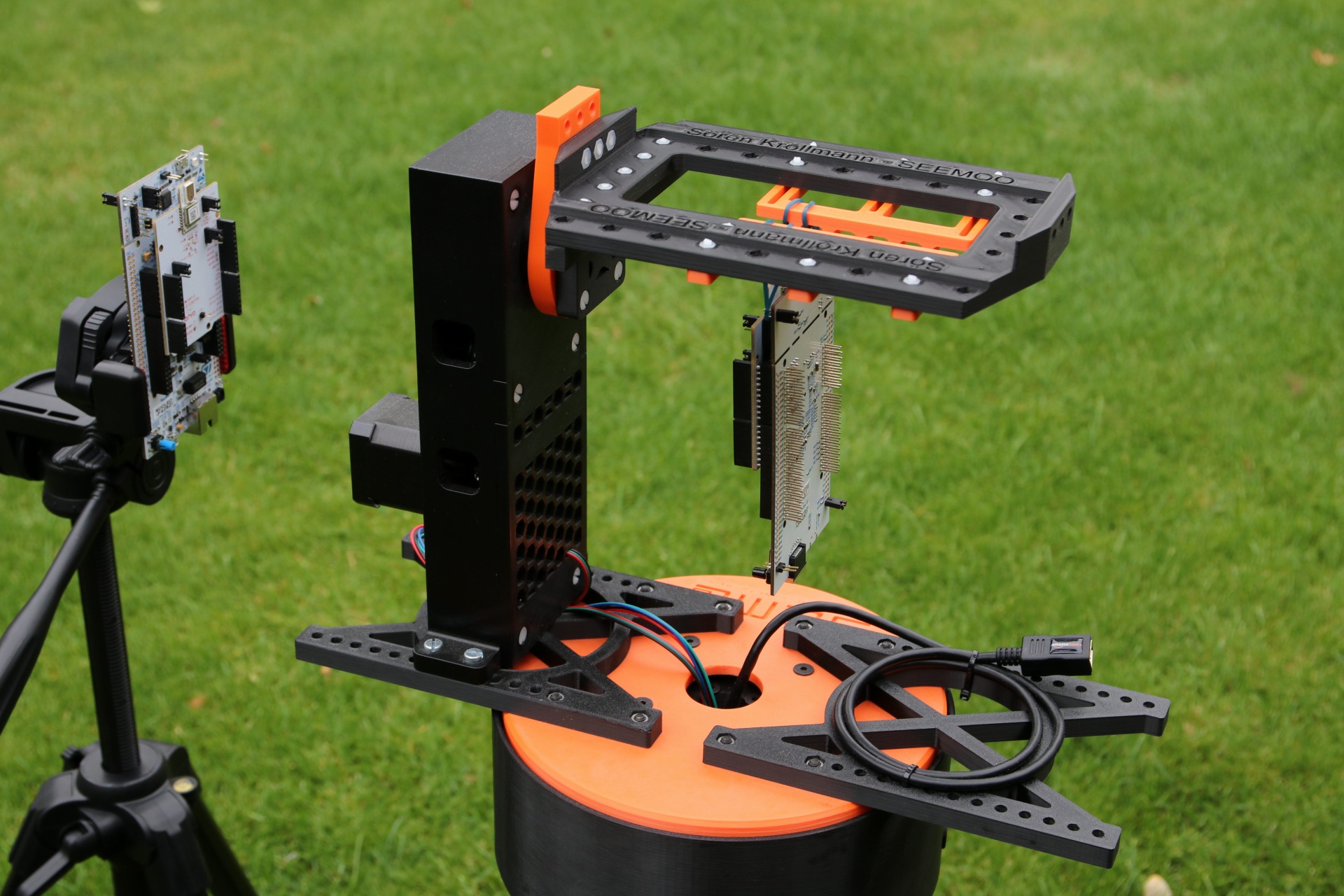}
    \caption{\raggedright{} NLOS measurement with the tower between the devices at a base rotation of $\theta=\SI{0}{\degree}$.}\label{fig:t2}
  \end{subfigure}
  \caption{Exemplary test setup for evaluating the influence of the tower on \ac{gwen}. The real installation also included the wiring, power supplies, and ensuring the correct spacing. This has been omitted in this picture to simplify matters.}\label{fig:gwen_arm_test}
\end{figure}

To evaluate the influence of \ac{gwen} on the recordings, we performed two tests using a pair of DW3000 \ac{uwb} devices. 
First, we placed both devices facing each other with no obstructions between them in clear \ac{los} (see \cref{fig:t1}). 
Second, we rotated the base of \ac{gwen} to $\theta=\SI{180}{\degree}$ such that the tower is now directly between both devices (see \cref{fig:t2}). We also turned the DW3000 again to ensure that they have the same orientation as in the first test.  For both tests, we ensured that the distance between the DW3000 \ac{uwb} devices is equal. 

In each test, we performed $10,000$ \ac{uwb} measurements. 
To assess the influence of \ac{gwen}, we calculated the mean measured distance and the standard deviation for both measurements.
In \ac{los}, without the tower, a mean value of \SI{35}{\centi\metre} with a standard deviation of \SI{1.3}{\centi\metre} was measured. 
With the tower, \SI{36}{\centi\metre} and \SI{1.7}{\centi\metre} were determined, respectively.

It is evident that the presence of the arm has increased the distance by \SI{1}{\centi\metre}.
Also, the standard deviation increased by \SI{0.4}{\centi\metre}. The expected accuracy and precision of these devices are claimed to be \SI{10}{\centi\metre}~\cite{qorvoinc.DWM3000Qorvo,qorvoinc.DW3000DataSheet2020} and therefore \ac{gwen} does not add a large enough error to disturb our measurements.

\subsection{Reproducibility}
We assess the reproducibility of measurements by using an experimental setup with two DW$3000$.
One board was our \ac{dut} mounted to \ac{gwen} and the other board was mounted on a tripod. Both were placed \SI{50}{\centi\metre} apart. 
\ac{gwen} rotated a full \SI{360}{\degree} at the base ($\theta$) and \SI{180}{\degree} at the arm ($\phi$) at each base position. At each position, the DW$3000$s performed $100$ \ac{uwb} distance measurements.

\begin{figure}[!t]
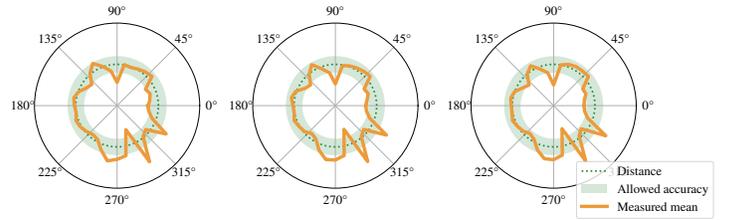

    \centering
    \begin{subfigure}{.16\textwidth}
      \centering
      \scalebox{0.32}{\input{figures/eval/gwen/polar_90_T1.pgf}}
      \caption{Recording 1}\label{fig:m1}
    \end{subfigure}%
    \begin{subfigure}{.16\textwidth}
      \centering
      \scalebox{0.32}{\input{figures/eval/gwen/polar_90_T2.pgf}}
      \caption{Recording 2}\label{fig:m2}\end{subfigure}%
    \begin{subfigure}{.16\textwidth}
      \centering
      \scalebox{0.32}{\input{figures/eval/gwen/polar_90_T3.pgf}}
      \caption{Recording 3}\label{fig:m3}
    \end{subfigure}
    \caption{Comparison of three 360° measurements made directly one after the other without changing the setup. Arm at 90°. The arm angle of $\phi=$\SI{90}{\degree} was chosen arbitrarily, other arm angles led to similar results.}\label{fig:m_comp}
\end{figure}

The same full measurement cycle was then performed three times in succession. 
Subsequent examination of the three records showed that the mean and standard deviation for all records were \SI{0.48}{\meter} and \SI{0.11}{\meter}, respectively.
\Cref{fig:m_comp} shows for each measurement the plot for a full \SI{360}{\degree} rotation of the base $\theta$ with a fixed arm angle of $\phi=\SI{90}{\degree}$.
All graphs have very similar shapes, and we can see similar results when comparing different arm angles. 

\begin{figure}[!t]
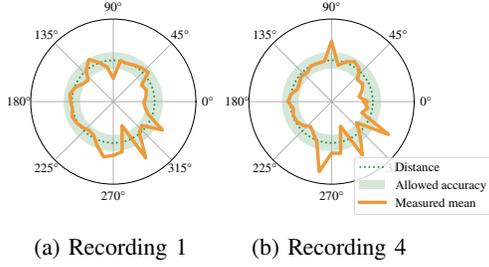

  \centering
  \begin{subfigure}{.16\textwidth}
    \centering
    \scalebox{0.32}{\input{figures/eval/gwen/polar_90_T1.pgf}}
    \caption{Recording 1}\label{fig:r1}
  \end{subfigure}%
  \begin{subfigure}{.16\textwidth}
    \centering
    \scalebox{0.32}{\input{figures/eval/gwen/polar_90_T4.pgf}}
    \caption{Recording 4}\label{fig:r4}
  \end{subfigure}
  \caption{Comparison of two \SI{360}{\degree} base rotation ($\theta$) measurements made one week apart. With arm rotation at $\phi=\SI{90}{\degree}$.}\label{fig:m_comp_2}
\end{figure}

Finally, we tore down the measurement setup and set it up again one week later. We measured with the same parameters and compared the results. 
\Cref{fig:m_comp_2} shows two plots one from the first recording described above and one from a fourth recording made one week later.
\Cref{fig:r1} and \Cref{fig:r4} both show a \SI{360}{\degree} rotation of the base ($\theta$) with the arm at $\phi=\SI{90}{\degree}$.
The general shape of the two measurements is similar, but we can see some deviations, which could have been caused by minor changes in the environment. 
The mean value is \SI{0.47}{\centi\metre}, \SI{1}{\centi\metre} less than the first measurements, but the standard deviation remained the same.

All tests combined show that \ac{gwen} only marginally influences \ac{uwb} measurements and can therefore be used to evaluate \ac{uwb} capable devices. 

\section{Experimental setup}\label{sec:experimental_setup}
\noindent 
In this section, we detail how we set up our measurement device \ac{gwen}, how we configured each \ac{dut}, and how we were able to collect distance measurement data from \ac{uwb} over a USB connection. 

We evaluated each device listed in \cref{tab:available_uwb} in three different environments to see how multipath effects and reflections might change the results.

\subsection{Measurement Setup}\label{ssec:general_setup}

\begin{figure}[!t]
    \centering
    \input{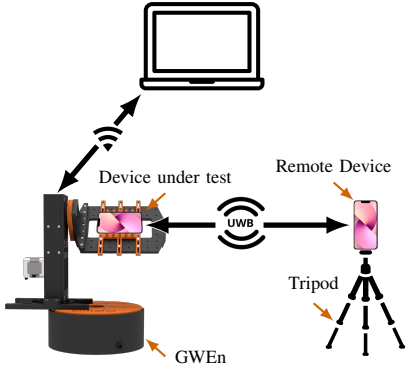}
    \caption{Setup of GWEn with a control PC and a remote device.}\label{fig:GWEn_setup}
\end{figure}

\begin{figure}[!t]
    \centering
    \begin{tikzpicture}
        
        \node [
            above right,
            inner sep=0] (image) at (0,0) {\includegraphics[width=.4\textwidth]{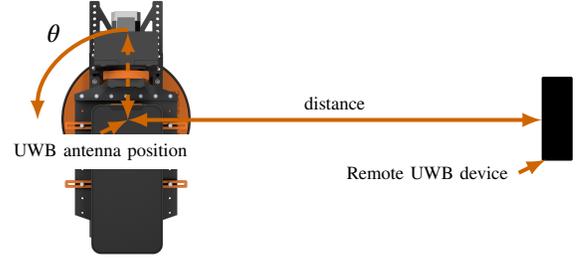}};
         
        \begin{scope}[
            x={($0.1*(image.south east)$)},
            y={($0.1*(image.north west)$)}]

            \tikzset{every path/.style={line width=1.5pt}};

            \coordinate (d) at (0.5,5.2);
            
            


            \draw[latex-, seemoo_organge] (9.5,3.7) -- ++(-0.5,-0.5)
                node[left,black,fill=white]{\scriptsize Remote UWB device};
            \draw[latex-, seemoo_organge] (1.85,5.2) -- ++(-0.5,-0.5)
                node[below,black,fill=white]{\scriptsize UWB antenna position};

            \draw [latex-latex, seemoo_organge]
            (9.42,5.2) coordinate (a) -- (1.85,5.2) coordinate (b) node[midway,above, black] {\scriptsize distance};
            \draw[latex-latex, seemoo_organge, dashed] (b) -- (1.85,8.5) coordinate (c);
            
            \pic[draw=seemoo_organge, -latex, angle eccentricity=1.2, angle radius=1.2cm] {angle=c--b--d};

            \node[] at (0.5,8.5) {$\theta$};
            
        \end{scope} 
    \end{tikzpicture}
    \caption{Top view of GWEn' setup that shows the base start position of $\theta=\SI{0}{\degree}$. In this position the arm is rotated to $\phi=0^\circ$.}\label{fig:GWEn_setup_top_view}
\end{figure}

\Cref{fig:GWEn_setup} shows the hardware setup for a typical measurement.
The setup consists of \ac{gwen}, two devices to test, a laptop, and a tripod.
The \ac{dut} is mounted on \ac{gwen}, which is placed on a camera tripod such that the \ac{dut} remains at a height of \SI{80}{\centi\metre}. This height should model the distance of a jeans' pockets to the ground.  
The \ac{dut} has been placed such that its antennas were in the intersection of the rotation axis of \ac{gwen} (see \cref{fig:GWEn_setup_top_view}).
The remote device is mounted on a tripod and a desired distance between them is set. 
The distance between both devices is measured using a laser meter or a measurement tape. 
The arm is set to $\phi=\SI{0}{\degree}$ by using an internal sensor that detects when the motor has rotated to the correct position. The base is aligned such that the tower is located at an angle of \SI{90}{\degree} as shown in \cref{fig:GWEn_setup_top_view}, which we define as $\theta=\SI{0}{\degree}$ (see \cref{fig:GWEn_setup_top_view}).

The rotation axes of \ac{gwen} and according to labels are plotted in \cref{fig:gwen_rotation_axes} and detailed in \cref{sec:gwen}.
For all our experiments \ac{gwen} performs these steps: 
\begin{enumerate}
    \item Start at arm rotation $\phi=\SI{0}{\degree}$ and base rotation $\theta=\SI{0}{\degree}$
    \item Rotate the arm $\phi$ by \SI{10}{\degree} 
    \item When the arm reaches $\phi=\SI{180}{\degree}$, rotate the base $\theta$ by \SI{10}{\degree}
    \item Rotate the arm $\phi$ by \SI{-10}{\degree} 
    \item When the arm reaches $\phi=\SI{0}{\degree}$, rotate the base $\theta$ by \SI{10}{\degree}
    \item Continue with step 2 until the base reaches $\theta=\SI{360}{\degree}$
\end{enumerate}

At each position, we measure the distance and collect the measurements, before \ac{gwen} continues with the next step. 

Each measurement for each pair of devices has been conducted in the same way in the same environments to ensure comparability.
Once the measurement is started all operations for the measurement happen on \ac{gwen} and the external laptop is no longer needed. For \ac{uwb} measurements, the ranging has to be started on the \ac{dut} and the remote device in advance. 

\begin{figure}[!t]
    \includegraphics[width=\linewidth]{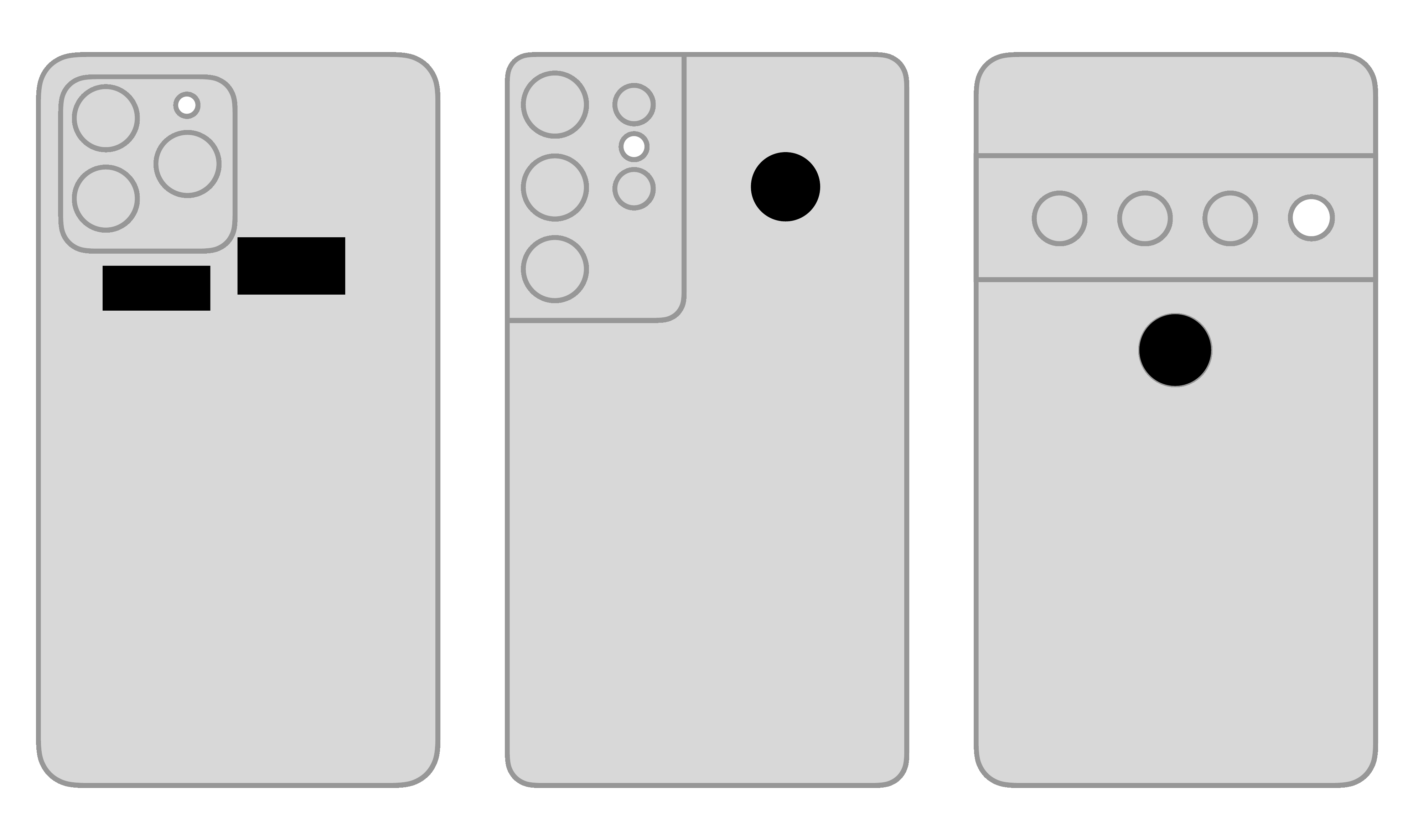}
    \caption{Left to right: Apple iPhone 12 Pro, Samsung Galaxy S21 Ultra, and Google Pixel 6. Pro. The \ac{uwb} antenna locations are marked in black.}\label{fig:smartphone_antenna_placement}
\end{figure}

\begin{figure}[!t]
    \centering 
    \includegraphics[width=\linewidth]{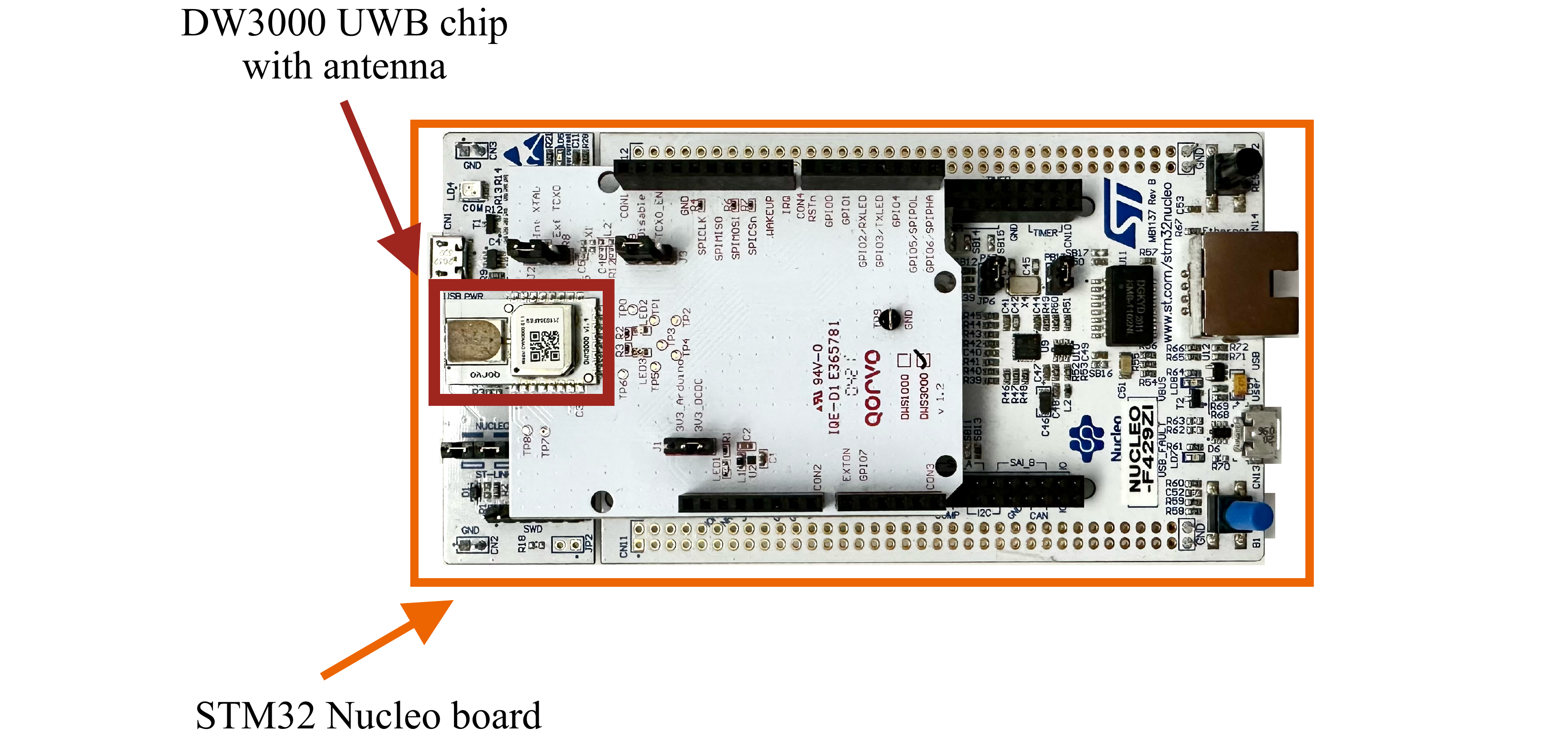}
    \caption{A DW3000EVB attached to a STM32 Nucleo board.}\label{fig:dw3000_annotated}
\end{figure}

\paragraph*{Antenna locations}
We identified the location of the \ac{uwb} transmission antenna in smartphones by using an oscilloscope and manually searching for the location with the strongest signal.
All manufacturers placed the antennas close to the rear cameras.
Since they all offer multiple receive antennas to determine the \ac{aoa} the receive antennas are often next to the transmission antenna~\cite{amaldevUWBTechApple2021}. We validated our claims by closely inspecting online device tear-downs~\cite{amaldevUWBTechApple2021,ifixitSamsungGalaxyS212021,dixonIPhone12122020,hughjeffreysPixelProTeardown2021}. All \ac{uwb} antennas are marked in \cref{fig:smartphone_antenna_placement,fig:dw3000_annotated}. 

\subsection{Evaluation environments}

\begin{figure*}
    \centering
    \begin{subfigure}[t]{0.32\textwidth}
        \centering
        \includegraphics[width=\textwidth]{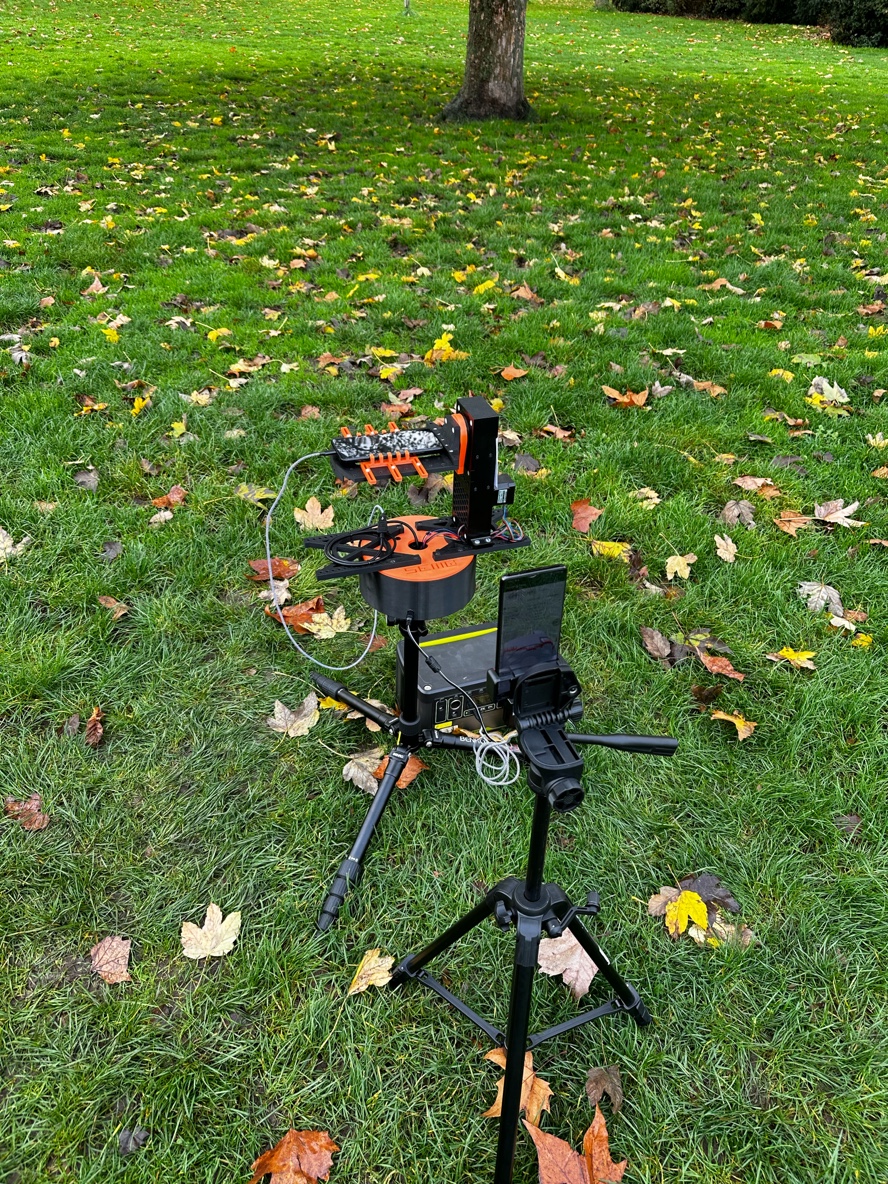}
        \caption{Outside}\label{fig:outside_picture}
    \end{subfigure}
    \hfill
    \begin{subfigure}[t]{0.32\textwidth}
        \centering
        \includegraphics[width=\textwidth]{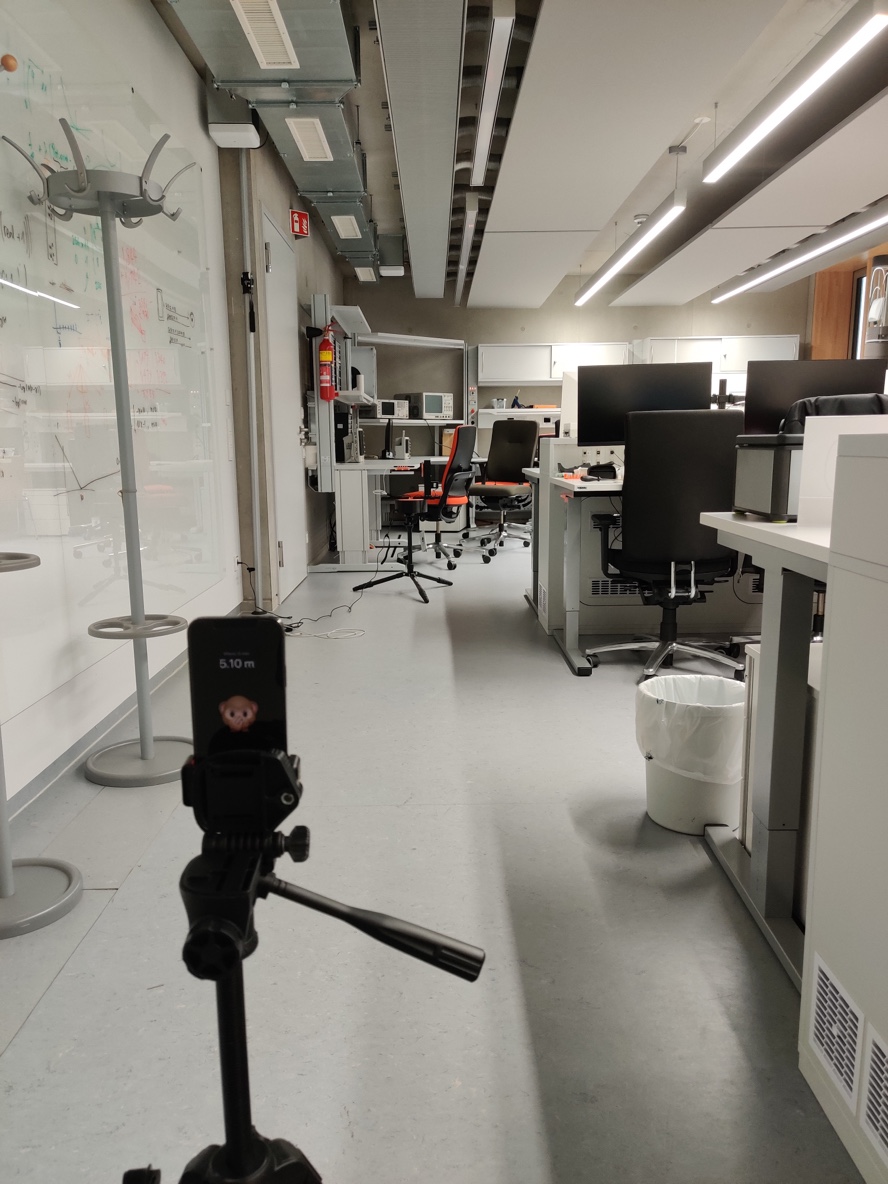}
        \caption{Lab}\label{fig:lab_picture}
    \end{subfigure}
    \hfill
    \begin{subfigure}[t]{0.32\textwidth}
        \centering
        \includegraphics[width=\textwidth]{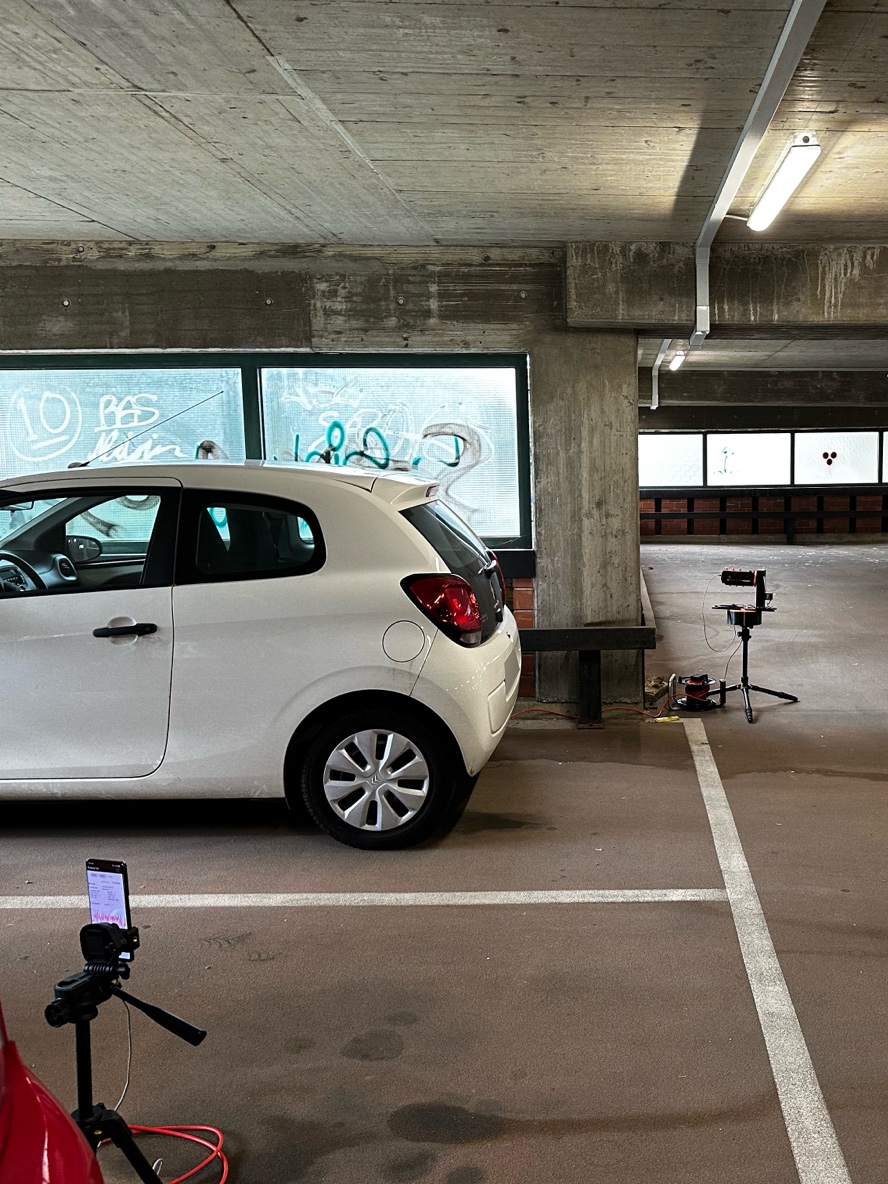}
        \caption{Garage}\label{fig:garage_picture}
    \end{subfigure}
    \hfill

    \caption{Our three measurement environments.}\label{fig:environment_pictures}

\end{figure*}

We decided on three environments: outside, our lab, and a parking garage. A photo of each environment is shown in \cref{fig:environment_pictures}.  

\paragraph{Outside}
A location outside with no reflecting objects nearby is a good environment to measure a ground truth with a low amount of multipath interference. The soil-based ground cannot reflect signals well. 
There were no nearby Wi-Fi signals that could have interfered with our measurements.
The only nearby Wi-Fi signal was emitted by \ac{gwen} itself and limited to the \SI{2.4}{\giga \hertz} band, which does not interfere with our tested \ac{uwb} channels. The setup is similar to the one used by~\cite{malajnerUWBRangingAccuracy2015,jimenezComparingDecawaveBespoon2016}. 

\paragraph{Lab} 
Our lab offers a range of difficulties for \ac{uwb} measurements that can also occur in an office environment. Our measurement location was close to a glass whiteboard, several computers, monitors, and office furniture. 
These office environments are likely locations for future application of \ac{uwb}--enabled locks that can replace the need for special key fobs. 

\paragraph{Public parking garage}
A \ac{uwb}-based \ac{pke} system is already deployed in new car models. A parking garage is a natural environment to evaluate this. 
Nearby cars are reflecting \ac{uwb} pulses and might cause collisions between them. Identifying the correct first path might become difficult as demonstrated in~\cite{279984,singhSecurityAnalysisIEEE2021a}. 

\subsection{Device configuration}\label{ssec:dev_conf}
All devices need to be configured in a way that they start measuring the distance using \ac{uwb}. We, therefore, explain the different configuration options and how we initiated the measurements. 
As \ac{uwb} is a rather new technology in consumer devices, there is no fully open \ac{api} available by any smartphone manufacturer, which would have allowed fine-grained configuration of the \ac{uwb} chip. Therefore, the configuration of \ac{uwb} parameters, like channel, preamble, data rate and \ac{sts} may not be changed. In many cases, we were also not able to extract the used parameters after the measurement. 

\paragraph{Apple}
The Apple \textit{NearbyInteraction} framework~\cite{appleinc.NearbyInteractionApplea} can be used by any iOS app and can be configured to measure either the distance to another iPhone, Apple Watch or to a compatible \ac{uwb} third-party chip. We use an \textbf{iPhone 12 Pro} as the \ac{dut} and an \textbf{iPhone 12 mini} as the remote device. Both devices utilize the same generation of Apple's U1 chipset. 

\paragraph{Samsung}
Samsung and Google both implement the \ac{fira} protocol and would be theoretically compatible with the other chipsets and the iPhone. Unfortunately, no documentation on how to initialize ranging with any other device is not available publicly. 

The Samsung Galaxy S21 Ultra that we used for testing comes with an installed \textit{UWBTest} app. This app is hidden and can only be launched by opening the telephone app and typing \texttt{*\#UWBTEST\#} in the phone number field. 
We use this app to run our \ac{uwb} measurements between two Samsung devices because this has been the most reliable option on Android. 
We cannot determine which channel or preamble is used here. A \textbf{Samsung Galaxy S21 Ultras} is used as the \ac{dut} and the remote device.

\paragraph{Google}
So far, there are only the Google Pixel 6 Pro and Pixel 7 Pro available that integrate a \ac{uwb} chip. The only supported feature with \ac{uwb} is Android Nearby Share~\cite{googleinc.HowWeRe2022} and the Android 12 \ac{uwb} API~\cite{googleinc.UltrawidebandUWBCommunication}. 
We developed a small test application that is able to perform \ac{uwb} ranging between compatible Android devices. Unfortunately, this \ac{uwb} API needs up to $10$ and may fail to measure the distance completely. All our measurements are conducted using Android 12. 
We configured the app to use channel 9 and the preamble code 11 for all our measurements. 
A \textbf{Google Pixel 6 Pro} is used as the \ac{dut} and the remote device. Additionally, we also performed all measurements with a \textbf{Samsung Galaxy S21 Ultra} as the remote device. Both measurements are compared in \cref{ssec:different_remote_device}.

\paragraph{Qorvo}
The Qorvo DWM3000EVB is a board that can be attached to an STM development board. By using the SDK provided by Qorvo, it's possible to write programs that can execute simple ranging measurements. Their sample code already provides most features needed for our evaluation. We configured the devices to use a static \ac{sts} and perform \ac{ds-twr} on channel 9 using preamble code 11. 

\subsection{Measurement result extraction}

Every device that we researched is able to generate results for the measurements and accessing them works differently on every device. As introduced in \cref{sec:gwen}, \ac{gwen} connects to all devices via USB to record measurement data while performing the measurement. We implemented separate parsers for each device that we support. 
If possible, we use raw measurements which are not enhanced by optimization algorithms. 

\paragraph{Apple} Apple devices often use extensive logging but omit sensitive data from the logs. To allow developers to debug their apps, certain logs can be activated using a \textit{debug profile}. We use the \textit{AirTag debug profile} to enable rich logs from \texttt{nearbyd}, a process which handles all \ac{uwb} related tasks~\cite{appleinc.ProfilesLogsBug}. 
With a USB connection to \ac{gwen}, we can fetch the logs and extract relevant measurement data. 
We extract raw measurements, as user-facing distance measurements are enhanced by custom machine learning algorithms.

\paragraph{Samsung} Samsung's devices also offer the option to increase the log verbosity by activating \textit{verbose vendor logging} in the developer settings. Then the smartphone logs all distance measurement data in a raw measurement and a calibrated measurement. Depending on the internal state the calibrated measurement may not be available.
\Ac{gwen} extracts both measurements using the \texttt{adb logcat} command line interface. 

\paragraph{Google} Google's smartphones log the \ac{uwb} distance measurements by default. Unfortunately, the logs are less verbose than Apple's or Samsung's. We, therefore, decide to use a custom logging format.  
We expect that all distance measurements on Google smartphones are raw measurements. Furthermore, we did not find any hints of a specially calibrated result or any machine learning-based enhancements. 

\paragraph{Qorvo} We mounted the DWM3000EVB to an STM32 development board. By programming the STM32 board, we are able to create a serial connection between the board and \ac{gwen}. Two-way communication allows \ac{gwen} to instruct the board to perform new measurements and to receive measurement results from the board. All results are raw results without any calibration. We tested the effect of calibrating the chips briefly, but we could not identify larger differences if compared to not calibrated devices.  


\definecolor{DW3000}{HTML}{293462}
\definecolor{Pixel}{HTML}{F24C4C}
\definecolor{Galaxy}{HTML}{EC9B3B}
\definecolor{iPhone}{HTML}{F7D716}
\newcommand\polarscale{0.66}
\newcommand\linescale{1.0}
\newcommand\subfigtextwidth{0.32\textwidth}

\ifsubmission{}
\else
\printinunitsof{in}\prntlen\textwidth
\fi

\section{Measurement Results}\label{sec:results}
\noindent 
In total, we conducted $28$ measurement series at three environments using four different devices. 
All \ac{uwb} capable smartphones have been evaluated at two different distances: \SI{5}{\metre} and \SI{0.5}{\metre}. 
We conducted the same measurements using the Qorvo DWM3000EVB as a reference. 
Measurements with larger distances were mostly not possible, since all smartphones fail to measure if the antennas are not directed at each other. 
All \acp{dut} have been placed on \ac{gwen} as described in \cref{sec:experimental_setup}. 
\ac{gwen} performed a \SI{360}{\degree} base rotation ($\theta$) and a \SI{180}{\degree} arm rotation ($\phi$) using a step size of \SI{10}{\degree}. At each position, a minimum of $10$ distance measurements have been recorded.
In this section, we state our measurement results, draw general statistics about each \ac{dut}, and take a look at each location to compare how devices behave differently depending on their environment.  

\subsection{Maximum distance}
In an initial measurement, we evaluated the maximum measurable distance of all smartphones in our test set. We used an outside location that has no obstructions.
The results are listed in \cref{tab:max_distance_uwb}. The iPhone is able to continuously measure the longest distance. We identified that the maximum is artificially limited to about \SI{40}{\meter} on the iPhone and \SI{23}{\meter} on the Samsung Galaxy S21 Ultra. In our evaluation, the measurements stopped at exactly that distance. 
The Google Pixel performed the worst here, and it was not possible to get reliable measurements at larger distances. The distance does not seem to be limited artificially.

\begin{table}
    \centering
    \caption{Maximum distance possible to measure with \ac{uwb}-enabled smartphones.}\label{tab:max_distance_uwb}
    \begin{tabular}[]{lr}
        \toprule
        Smartphone & Distance (m) \\ 
        \midrule
        Apple iPhone 12 Pro & 40.0 \\
        Samsung Galaxy S21 Ultra & 23.0 \\ 
        Google Pixel 6 Pro & 11.6 \\
        \bottomrule
    \end{tabular}
\end{table}

\subsection{Overall results}

\begin{figure}[!t]
    \includegraphics[width=\columnwidth]{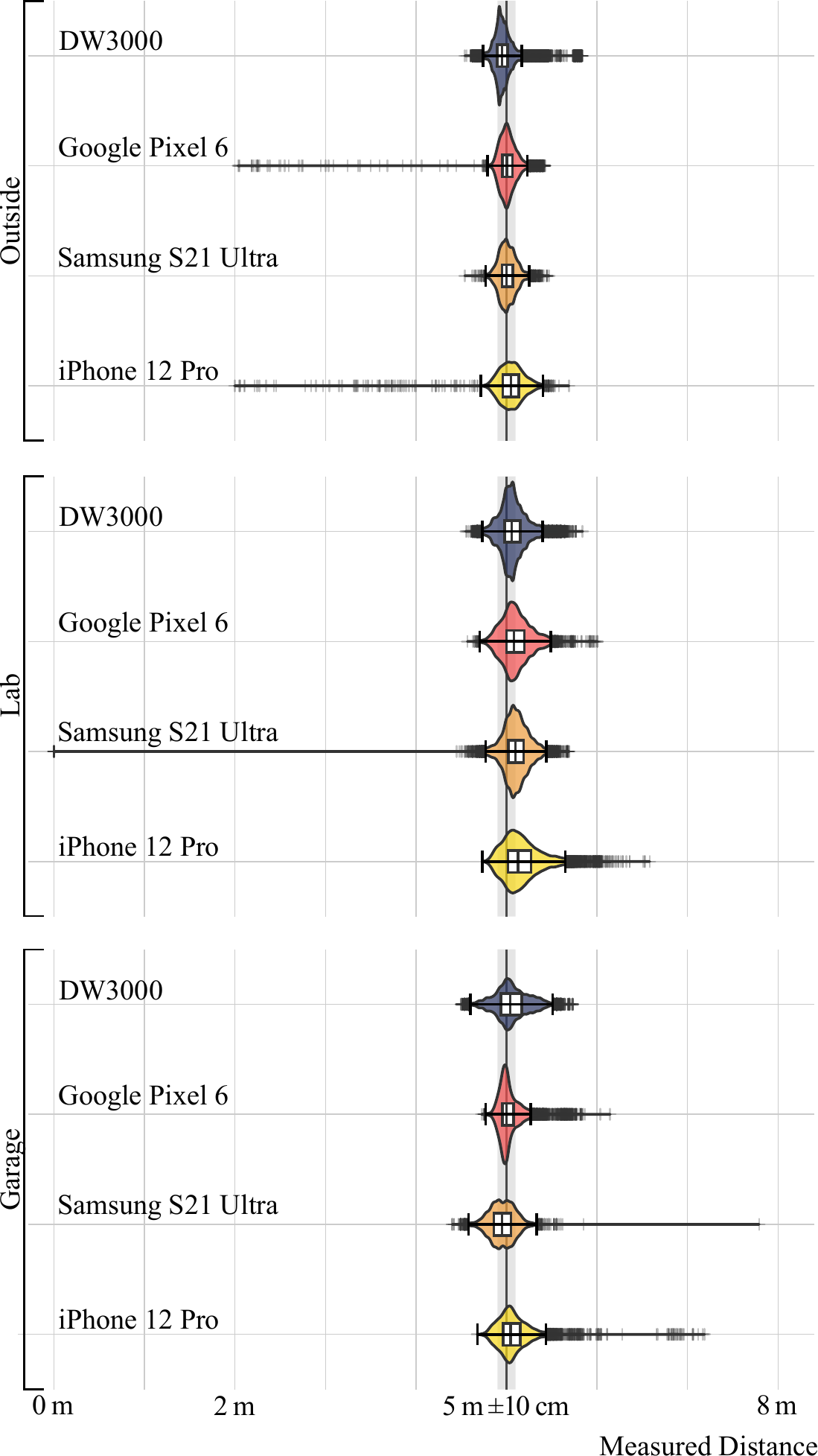}
    \caption{Distances measured outside, in a lab environment, and in a parking garage. The true distance (\SI{5}{\meter}) is marked in the plot along with the expected accuracy of $\pm$\SI{10}{\centi\metre}. The box plots show quartiles, median, and outliers. The violin plots show a kernel density estimate of the measurements.}\label{fig:violin}
\end{figure}

\begin{figure*}
    \captionsetup[subfigure]{justification=centering}
    \centering
    \begin{subfigure}[t]{\subfigtextwidth}
        \centering
        \includegraphics{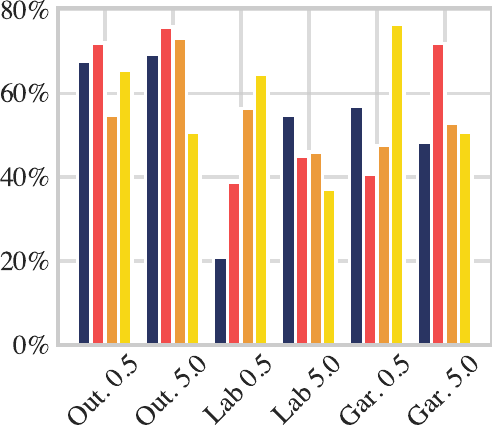}
        \caption{Accuracy}\label{fig:accuracy}
    \end{subfigure}
    \hfill
    \begin{subfigure}[t]{\subfigtextwidth}
        \centering
        \includegraphics{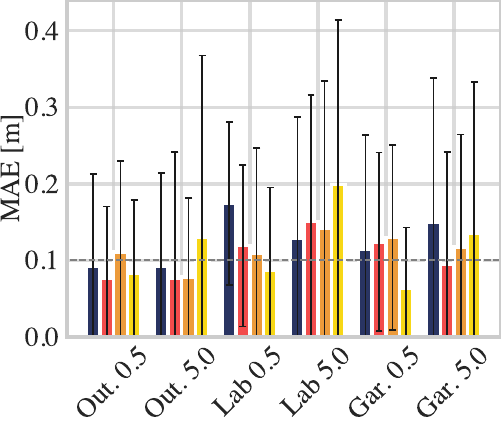}
        \caption{Mean absolute error (MAE) with standard deviation (SD)}\label{fig:mae}
    \end{subfigure}
    \hfill
    \begin{subfigure}[t]{\subfigtextwidth}
        \centering
        \includegraphics{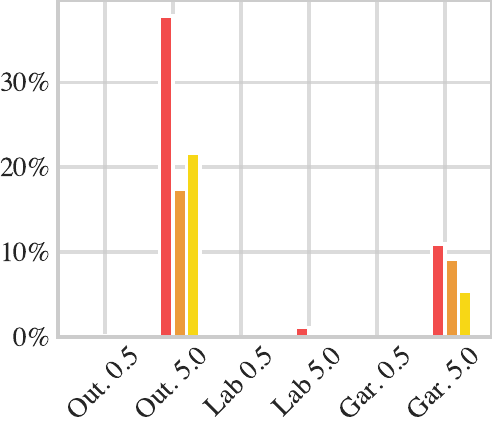}
        \caption{Failed measurement positions}\label{fig:failed_measurements}
    \end{subfigure}
    \hfill

    \caption{ Statistical data of all measurements performed on the
         {\color{DW3000} \rule{3pt}{9pt}  DW3000}, {\color{Pixel} \rule{3pt}{9pt} Google Pixel 6 }, {\color{Galaxy} \rule{3pt}{9pt} Samsung S21 Ultra}, and {\color{iPhone} \rule{3pt}{9pt} iPhone 12 Pro}. Failed measurements are excluded. }

\end{figure*}

\Cref{fig:violin} plots the results of all measurements conducted at a \SI{5}{\metre} distance. The plot is a combination of a violin plot and a box plot. 
We can see that most measurements are in proximity to the expected distance. 
Interestingly, the \acp{dut} have shown large maximum errors. The Samsung Galaxy reduced the distance by up to \SI{5}{\meter}, the Apple iPhone, and the Google Pixel by up to \SI{3}{\meter}. 

To evaluate the accuracy we define a range of $\pm$\SI{10}{\centi\meter} the actual distance as accurate. An error of $\pm$\SI{10}{\centi\meter} is advertised by chip manufacturers~\cite{qorvoinc.DW3000DataSheet2020,amosenseco.ltdAmotechSR040Module2021}. 
Since the antenna is in the intersection of the rotation axes, movements of \ac{gwen} did not alter the distance. In \cref{fig:violin} the accuracy range is marked in gray. 
\Cref{fig:accuracy} shows the percentage of accurate measurements. The accuracy ranges between $20\%$ and $60\%$. Depending on the environment and distance, for a given device it can vary by up to $50\text{pp}$. 

Although the maximum errors can be large, \cref{fig:mae} shows that the \ac{mae} is less than \SI{20}{\centi\metre}, while the mean \ac{sd} is below \SI{25}{\centi\metre} for each measurement series. \Cref{fig:failed_measurements} plots the percentage of failed measurements. A failed measurement is a position ($\theta$ \& $\phi$) of \ac{gwen} at which less than $10$ measurements were recorded in $\SI{30}{\second}$. A high failure rate translates into an unreliable \ac{uwb} system. 

\begin{figure*}[!t]
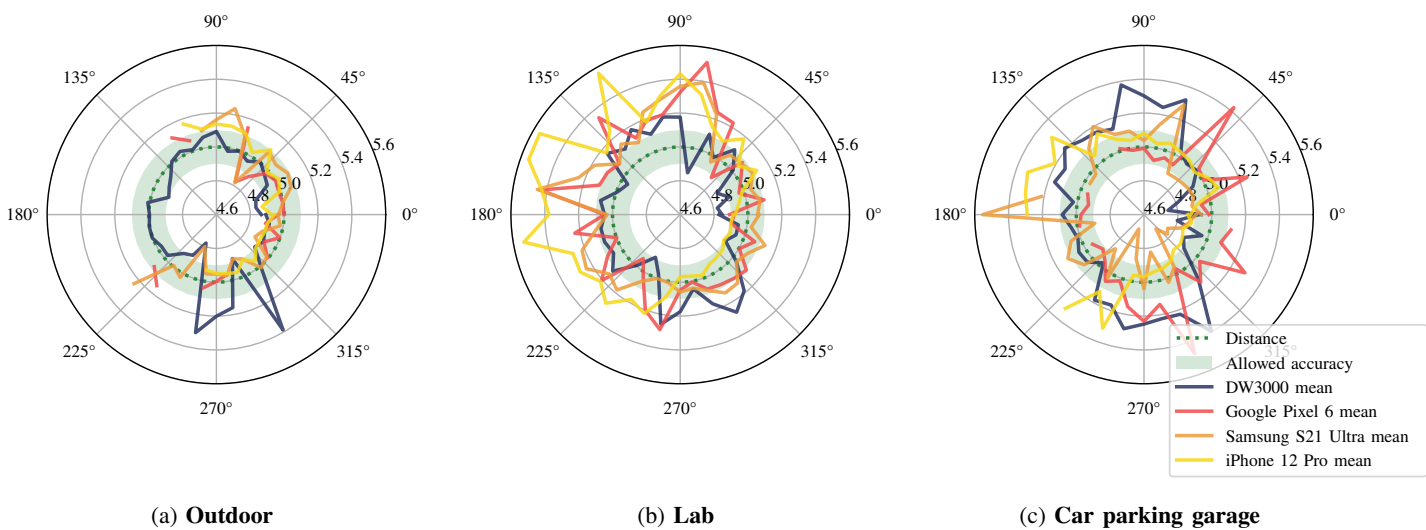

    \captionsetup[subfigure]{justification=centering}
    \centering

    \begin{subfigure}[t]{\subfigtextwidth}
        \centering 
        \scalebox{\polarscale}{\input{figures/eval/devices/polar_plots/polar_90_Outside_5.pgf}}
        \caption{\textbf{Outdoor}}\label{fig:outdoor_polar}
    \end{subfigure}
    \hfill 
    \begin{subfigure}[t]\subfigtextwidth
        \centering 
        \scalebox{\polarscale}{\input{figures/eval/devices/polar_plots/polar_90_Lab_5.pgf}}
        \caption{\textbf{Lab}}\label{fig:lab_polar}
    \end{subfigure}
    \hfill 
    \begin{subfigure}[t]\subfigtextwidth
        \centering 
        \scalebox{\polarscale}{\input{figures/eval/devices/polar_plots/polar_90_Car_5.pgf}}
        \caption{\textbf{Car parking garage}}\label{fig:car_polar}
    \end{subfigure}

    \caption{Mean Measurements at $\SI{5}{\meter}$ distance with one full base rotation ($\theta \in [\SI{0}{\degree},\SI{350}{\degree}]$). Arm position is fixed to $\phi=\SI{90}{\degree}$}\label{fig:polar_all}

\end{figure*}

\Cref{fig:polar_all} shows a polar plot of a full base rotation ($\theta \in [\SI{0}{\degree},\SI{350}{\degree}]$) with the arm position fixed at $\phi = \SI{90}{\degree}$. The remote device is placed at the angle of $\theta = \SI{0}{\degree}$ at a distance of \SI{5}{\meter}. This plot corresponds to the setup shown in \cref{fig:GWEn_setup_top_view}.

\subsection{Outside}


This section covers our measurements conducted outside, on soil ground without any obstructions that could reflect the measurements. 
Missing surfaces to reflect the signals influence the measurements. At larger distances, the signals that are directed away from the remote device, do not reach it or the \ac{dut} cannot receive signals while the antenna is directed away from the remote device. 

\paragraph*{Accuracy} 
For most devices, the measurements outside resulted in the highest percentage of accurate measurements. This likely stems from the minimal interference and lack of obstacles that could have produced reflections. As visible in \cref{fig:outdoor_polar} there is a correlation between the base position and the accuracy. 
Overall the Google Pixel 6 Pro had the best accuracy, followed by the DW3000, which created clearly visible outliers at several positions. 

\paragraph*{Measurement error}
The \ac{mae} and \ac{sd} of most devices are low and close to the goal of \SI{10}{\centi\meter}. Only the iPhone has a largely increased \ac{sd} of \SI{23}{\centi\meter} at the \SI{5}{\meter} distance. All other devices produce similar results. 
Moreover, the Google Pixel and the iPhone also produced outliers of up to \SI{-3}{\meter}. Previous research~\cite{singhSecurityAnalysisIEEE2021a} expected these failures only in a multipath environment, but our measurements show that they can also occur without reflections. 

\paragraph*{Reliability}
Interestingly, we can see very large failure rates on all smartphones at a distance of \SI{5}{\meter}. By rotating the base ($\theta$) of \ac{gwen} the \ac{dut}'s antenna will point away from the remote device when $\theta \in [90, 270]$. The signals are no longer strong enough to reach both devices. The effect is the strongest, when the arm is at $\phi = \SI{90}{\degree}$ and the phone internals are shielding the \ac{uwb} antenna. 
In \Cref{fig:outdoor_polar} failed measurements are not plotted. 
The Google Pixel is the most unreliable here and fails to measure at \SI{37.8}{\percent} of all positions. 
At the \SI{0.5}{\meter} distance this effect was not visible. 
The DWM3000 has an exposed antenna and therefore does not fail to receive signals at any position.

\subsection{Lab}

In this section, we cover the results of the measurements conducted in our lab. This location had the most surfaces and reflecting areas, i.e., whiteboards and screens, in our measurement series.
This resulted in very good reliability but reduced accuracy. 

\paragraph*{Accuracy}
Depending on the distance and the device, the accuracy in this environment varies a lot. The DW3000 has an accuracy of \SI{20.8}{\percent} at \SI{0.5}{\meter}, while the Samsung reaches \SI{56.4}{\percent}. \Cref{fig:lab_polar} shows how the accuracy on all smartphones is worse when $\theta \in [90,270]$. Most outliers are distance enlargements due to longer paths caused by reflections of the signals. Only the Samsung Smartphone measured reduced distances of \SI{0.0}{\meter} and \SI{0.03}{\meter}. 

\paragraph*{Measurement error}
The Apple iPhone has the largest \ac{mae} of \SI{19.9}{\centi\meter}, which is also visible in \cref{fig:lab_polar}. The iPhone may select the wrong first path and therefore, increase or even decrease the measured distance. 
This behavior also aligns with a previously presented attack on Apple's U1 \ac{uwb} chip, which has shown that the iPhone can be tricked into accepting an attacker's pulse as the first path and, therefore, producing shortened distance measurements~\cite{279984}. 

However, this error is only present at larger distances. The same characteristic that blocked the signals outside, is now resulting in increased errors. 

\paragraph*{Reliability}
The strong multipath effects in this environment result in fewer failed measurements (see \cref{fig:failed_measurements}). The Google Pixel has the highest failure rate of \SI{1}{\percent} all other devices are able to measure reliably at all positions.
At \SI{0.5}{\meter} all devices measured reliably.

\subsection{Parking garage}

The public parking garage had over twenty cars on one level. The results of these measurements are presented in this section. The remote devices were placed next to a parked car. As always, both devices were placed in \ac{los}. Multipath effects occur in this environment, but they are less common than in the lab environment. 

\paragraph*{Accuracy}
The overall accuracy of all smartphones increased compared to the lab environment. The Google Pixel performs best at \SI{5}{\meter} reaching an accuracy of \SI{71.9}{\percent}. All other devices performed similarly and were accurate in \SI{50}{\percent} of the measurements. For \SI{0.5}{\meter} the iPhone is the most accurate with \SI{76.5}{\percent} of accurate measurements. 

\paragraph*{Measurement error}
The measurement error of all devices is less than \SI{14}{\centi\meter} in an environment where this technology is already deployed in a \ac{pke} system for modern cars. Additionally, we did not measure large negative outliers on any device. 
All smartphones produced distance enlargements and Samsung Galaxy measured a distance of up to \SI{+2.79}{\meter}. 

\paragraph*{Reliability}
The reliability in this environment is good, with less than \SI{10}{\percent} of failed measurements on any device and distance. Since the garage had a concrete roof, concrete floor and many cars around, the reflections of the \ac{uwb} signal were able to reach the remote device.
Similar to the outside environment, the smartphone internals block the signal transmission and the environment did not provide enough reflections to allow for multipath transmissions. 
In practice, this means, that smartphones might need several attempts to measure the distance to a car from a larger distance, but should not fail.
 
In \cref{fig:car_polar} we can see fewer failed ranging measurements as in the outdoor environment. 
At base rotations of  $\theta \in [\SI{100}{\degree}, \SI{230}{\degree}]$, several measurements failed. Similarly to the outside environment, the smartphone internals block the signal transmission and the environment did not provide enough reflections to allow for multipath transmissions.

\subsection{Influence of the remote device}\label{ssec:different_remote_device}

As mentioned in \cref{ssec:dev_conf}, the iPhone and Android smartphones are not compatible to perform \ac{uwb} distance ranging with each other. Fortunately, Android devices from different manufacturers are compatible, and we, therefore, performed all measurements of the Google Pixel as a \ac{dut} also with a Samsung Galaxy as the remote device. 

Unfortunately, we were not able to measure with a Google Pixel as a remote device in the outdoor environment. This was supposed to be our last measurement series, but the phones were not able to connect at all over \ac{uwb}. A software reset of the devices did not result in any change. Therefore, our outside measurements of the Google Pixel purely use a Samsung Galaxy as the remote device. 

We measured a consistent improvement in all areas when using the Samsung Galaxy as the remote device. 
The measurements failed less often and the accuracy increased by $5\text{pp}$  to $10\text{pp}$. The \ac{mae} is slightly worse in this setting and outliers were similar. The remote device has an influence on the overall measurement performance. The Samsung Galaxy might have a stronger antenna, and therefore, reduces the failure rate.


\section{Discussion}\label{sec:discussion}
\noindent 
Our measurements have shown that the advertised accuracy of $\pm$\SI{10}{\centi\metre} 
is not reproducible. Even when both \ac{uwb} antennas were directed at each other a \ac{mae} of more than \SI{10}{\centi\metre} is normal. Additionally, all smartphones are unreliable in certain positions and environments. They failed to measure the distance at positions directed away from the remote device. 
However, the overall performance is promising: The \ac{mae} is below \SI{20}{\centi\meter} and at shorter distances, the devices succeeded to measure the distance at every position we evaluated. 
In this section, we detail how these results may affect \ac{uwb}-based entry systems and which considerations have to be taken when implementing such systems. 

\subsection{Overall performance}

\paragraph*{Accuracy and Reliability}
Our measurements identified that internal components of a smartphone shield the low power \ac{uwb} signals, resulting in strong directional differences. In a real-world scenario, this influences the overall user experience: Depending on the position of a smartphone in a pocket or bag, the device is not able to perform \ac{uwb} ranging. 
If successful, measurements can have an increased distance by several meters because the devices rely on reflected signals, which travel a longer path. 
These issues can delay the time until a distance-based action is triggered, e.g., when approaching \ac{uwb}--cars with iPhone support have the ability to activate their headlights in order to welcome the driver~\cite{incExploreUWBbasedCar}.
In addition, future smart home interactions might fail to recognize that the user is present and as a result turn off lights or heating in these areas. 
At short distances we measured a high reliability of all devices and a \ac{mae} of less than \SI{20}{\centi\meter}.

\paragraph*{Software support}
All smartphones, which we evaluated offer an API to perform \ac{uwb} ranging to other smartphones or \ac{iot} devices. Apple was the first manufacturer to open up their ranging system to \ac{fira} compatible chipsets (e.g., Qorvo and NXP)\cite{NearbyInteractionApple}. Google also integrated an Android-wide API for \ac{uwb}, which should also be compatible with \ac{fira} chips~\cite{googleinc.UltrawidebandUWBCommunication}. Unfortunately, the documentation lacks important parts and Google does not disclose how to perform ranging with non-Android devices. Moreover, the stability of the API varies largely. To start ranging successfully it was required to perform around ten attempts on all Android devices and finally, our Google Pixel devices suddenly failed to perform \ac{uwb} ranging completely. 

Overall, the user experience may be weakened, but the mean measured distances over ten or more consecutive measurements have proven to normalize the results. If the software issues are resolved, the current consumer-market implementations are in a good shape to be used for distance estimations in most smart home scenarios and for secure entry systems, if some precautions are made (see \cref{sec:discussion_recommendations}).

\subsection{Comparison to previous research}\label{ssec:comparison_related_work}


Several researchers analyzed the accuracy and ranging errors of different \ac{uwb} devices. None of these devices were actual consumer hardware, most of them focused on the Decawave (now Qorvo) devices. 
In this section, we compare the results of our research on consumer hardware and the Qorvo DW3000 with previous results. 


\paragraph{DW1000} Malajner et al.~\cite{malajnerUWBRangingAccuracy2015}  analyzed the ranging accuracy in \ac{los} of the DW1000, the predecessor of the DW3000. 
Depending on the configuration of the chip, they identified a \ac{rmse} between \SI{28.19}{\centi\meter} and \SI{54.04}{\centi\meter} for uncalibrated devices. For calibrated devices, the \ac{rmse} was lower at a range of \SIrange{4.34}{14.04}{\centi\meter}. 

The \ac{rmse} for an uncalibrated DW3000 over all our measurements was \SI{16.03}{\centi\meter}, which includes all positions and all environments. For comparison, the Google Pixel has a \ac{rmse} of \SI{14.98}{\centi\meter},the Samsung Galaxy S21 Ultra \SI{15.65}{\centi\meter}, and the iPhone \SI{17.68}{\centi\meter}. All of our results are in a similar range when compared to the older DW1000.


\paragraph{BeSpoon phone} Jimenez et al.~\cite{jimenezComparingDecawaveBespoon2016} compared the BeSpoon phone\footnote{The BeSpoon phone was a prototype hardware demonstrating that it is feasible to implement \ac{uwb} in smartphones.} with the DW1000.  
In \ac{los} measurements, the BeSpoon reached a mean error of \SI{2.6}{\centi\meter} and a \ac{sd} of \SI{11}{\centi\meter}. In \ac{nlos}, the mean error was \SI{50}{\centi\meter} and the \ac{sd} \SI{61}{\centi\meter}. Using the mean error is non-optimal as the errors can be positive or negative, which is why we measure the \ac{mae} in this work. When comparing the mean standard deviation over all our measurements, the DW3000 had \SI{14.03}{\centi\meter}, the Pixel \SI{13.04}{\centi\meter}, the Galaxy S21 Ultra \SI{13.63}{\centi\meter}, and the iPhone \SI{15.02}{\centi\meter}. We can see that modern consumer hardware operates at similar levels, as our measurements include both \ac{nlos} (blocked by smartphone internals) and \ac{los} scenarios. 


\paragraph{DW3000 and 3dB Access} Fluearto et al. \cite{flueratoruEnergyConsumptionRanging2020} compared the DW3000 with a \ac{uwb}--\ac{lrp} development kit from 3db Access. In \ac{los}, the mean error of the 3dB Access device was \SI{4.8}{\centi\meter} whereas the mean error of the DW3000 was \SI{4.1}{\centi\meter}. They measured \acp{sd} of \SI{8.4}{\centi\meter} and \SI{4.1}{\centi\meter}, respectively. In \ac{nlos}, the mean error increased to \SI{62.5}{\centi\meter} and \SI{34}{\centi\meter}. The \ac{sd} also increased to \SI{104}{\centi\meter} and \SI{35}{\centi\meter}. Our results are in the same range as the ones measured by Fluearto et al.

\paragraph{DW3000 in motion} Tiemann et al. \cite{tiemannExperimentalEvaluationIEEE2022a} performed an experimental evaluation of the DW3000 while walking through a storage hall with many obstructions. Overall, the measurement errors are less than \SI{50}{\centi\meter} throughout the whole measurement series.
In our measurement campaign, the DW3000 performed similarly well with only a few larger outliers.
Our measurements on consumer hardware, however, had larger outliers reaching multiple meters.

\subsection{Security implications}

The already mentioned shielding combined with reflections resulted in an enlarged measured distance in different environments. At a distance of \SI{5}{\metre}, the measured distances on all smartphones resulted in range of \SI{0}{\metre} to \SI{7.79}{\metre}. An enlargement of the measured distance can affect the convenience of a \ac{pke} system, but not the security. 

The Samsung devices show a lower \ac{mae} and better accuracy than the iPhones, but the Samsung devices repeatedly measured a distance of \SI{0}{\meter} in the Lab environment. Such strong reductions can affect the security of \ac{uwb}-based entry systems. 
The Apple iPhone and Google Pixel also created distance reductions of up to \SI{-3}{\meter} when ranging outside. This might not be enough to provoke the unlock action from a \SI{5}{\meter} distance. However, these reductions are unexpected, since the devices had no interference and should, therefore, be able to measure accurately.  

None of these distance reductions have been consistent and happened as outliers. Following our recommendations in \cref{sec:discussion_recommendations} mitigates these problems. 

Previous research presented an attack on iPhones in a \ac{los} setup that resulted in reproducible distance reductions~\cite{279984}. It is to be researched if the attack's success-rate increases when both devices are no longer in \ac{los}, use different orientations as used in our experimental setup, or operate in environments with strong interferences. 
Especially on the iPhone, we discovered that the \ac{sd} increases when the device is no longer directed to its peer (see \cref{fig:lab_polar}).

\subsection{Recommendations for implementing a secure passive keyless entry system}\label{sec:discussion_recommendations}

\ac{uwb} is the most promising technology when it comes to secure distance measurements at a consumer hardware level. Known attacks only have a low success rate and in general, the produced ranging errors are less than \SI{20}{\centi\metre}. Nevertheless, there are important things to consider when implementing such a system. 

Since this technology directly interferes with the real world by sending pulse-based signals, any obstructions, the positioning of the devices, and other wireless devices can influence the accuracy of distance measurements. This has been demonstrated by our study and previous work~\cite{tiemannExperimentalEvaluationIEEE2022a}.

To implement a secure entry system we give four recommendations: 
\subsubsection{Two-way ranging}
The implementation should always use the IEEE 802.15.4z \acf{ds-twr} ranging algorithm. With \ac{ds-twr} ranging, potential clock drifts are eliminated, and both sides calculate a distance measurement without the necessity to send the measured distance from one device to the other.
If a smartphone sends a command to \textit{open} to a smart lock, the measured distance should also be validated by the lock before granting access. 

\subsubsection{Use mean values}
Before a trusted distance can be estimated, at least ten to $15$ ranging cycles should be performed. 
Using a sliding window of ten measurements it is possible to calculate a mean distance over all measured distances. Only the mean value should be used for the decision of granting access or not. In a security context, the device should never perform the distance-based action on the first measurement that is in the desired range. 
Our study has shown that accidental distance reductions of \SI{3}{\meter} to \SI{5}{\meter} can happen. According to our tests, in most scenarios, ten measurements can be performed in less than \SI{2}{\second}. 

\subsubsection{Drop clearly wrong measurements}
When placing two devices right next to each other it is possible to create negative values for a distance measurement. These values are very rare and should not be considered in any parts of the algorithm. 

\subsubsection{Detect potential attacks}
The only currently known attack on \ac{uwb} \ac{hrp} IEEE 802.15.4z systems has a low success rate and cannot be controlled accurately~\cite{279984}. An iPhone under attack has reported several measurements with up to \SI{-2}{\meter}. Therefore, negative values should also be counted to detect potential attacks on an \ac{uwb} entry system. 
If reported distances fluctuate strongly or the system detects several strongly negative values, the distance measurement should be suspended, and the user should be informed. 


\section{Conclusion}\label{sec:conclusion}
\noindent 
In our study, we evaluated  \ac{uwb} distance measurements of three modern smartphones with \ac{uwb} and a development kit from Qorvo with a focus on accuracy and reliability. We designed, built and used a novel measurement setup called \ac{gwen} to evaluate changes in the antenna position on performed measurements.

Our results have shown that in most cases the measured distance is less accurate than advertised by the manufacturers\cite{qorvoinc.DW3000DataSheet2020,nxpsemiconductorsSR040UltraWidebandTransceiver2021}. Nevertheless, the devices produce errors in a reasonable range with a mean absolute error of less than \SI{20}{\centi\meter}. The results are on average good enough to perform distance estimation and to decide whether a smart lock or \ac{pkes} of a car should open. 
However, all smartphones shared the weakness that measurements can fail depending on the position of the antenna, e.g., when the antenna no longer faces the direction of the remote device. These issues decrease the user experience and may lower the adoption of people. 
From our results, we derived four recommendations to improve the security of \ac{pke} systems.

\section*{Acknowledgment}
\noindent This work has been funded by the German Federal Ministry of Education and Research and the Hessen State Ministry for Higher Education, Research and the Arts within their joint support of the National Research Center for Applied Cybersecurity ATHENE and by the LOEWE initiative (Hesse, Germany) within the emergenCITY center.

\bibliographystyle{IEEEtran} 
\bibliography{main.bib}

\begin{acronym}
    \acro{uwb}[UWB]{Ultra-Wide Band}
    \acro{hrp}[HRP]{High-Rate Pulse Repetition Frequency}
    \acro{pkes}[PKES]{Passive Keyless Entry and Start}
    \acro{pke}[PKE]{Passive Keyless Entry}
    \acro{lrp}[LRP]{Low-Rate Pulse Repetition Frequency}
    \acro{twr}[TWR]{Two-Way Ranging}
    \acro{ds-twr}[DS-TWR]{Double-Sided Two-Way Ranging}
    \acro{ss-twr}[SS-TWR]{Single-Sided Two-Way Ranging}
    \acro{tdoa}[TDOA]{Time Difference of Arrival}
    \acro{sfd}[SFD]{Start of Frame Delimiter}
    \acro{prf}[PRF]{Pulse-Repetition Frequency}
    \acro{psd}[PSD]{Power Spectral Density}
    \acro{toa}[ToA]{Time-of-Arrival}
    \acro{tof}[ToF]{Time-of-Flight}
    \acro{los}[LoS]{Line-of-Sight}
    \acro{nlos}[NLoS]{Non-Line-of-Sight}
    \acro{sts}[STS]{Scrambled Timestamp Sequence}
    \acro{sfd}[SFD]{Start-of-Frame Delimiter}
    \acro{mitm}[MitM]{Machine-in-the-Middle}
    \acro{ni}[NI]{Nearby Interaction}
    \acro{ble}[BLE]{Bluetooth Low Energy}
    \acro{edlc}[ED/LC]{Early Detect/Late Commit}
    \acro{bpsk}[BPSK]{Binary Phase Shift Keying}
    \acro{bpm}[BPM]{Burst-Position Modulation}
    \acro{rssi}[RSSI]{Received Signal Strength Indicator}
    \acro{gwen}[GWEn]{\textbf{G}imbal-based platform for \textbf{W}ireless \textbf{E}valuatio\textbf{n}}
    \acro{dut}[DUT]{device under test}
    \acro{ani}[ANI]{Apple Nearby Interaction}
    \acro{fira}[FiRa]{Fine Ranging Consortium}
    \acro{ccc}[CCC]{Car Connectivity Consortium}
    \acro{aoa}[AoA]{angle of arrival}
    \acro{api}[API]{application programming interface}
    \acro{pla}[PLA]{polylactic acid}
    \acro{cdf}[CDF]{cumulative Distribution Function}
    \acro{mae}[MAE]{mean absolute error}
    \acro{sd}[SD]{standard deviation}
    \acro{rks}[RKS]{Remote Keyless System}
    \acro{nfc}[NFC]{near-field communication}
    \acro{iot}[IoT]{Internet of Things}
    \acro{prf}[PRF]{pulse repetition frequency}
    \acro{rmse}[RMSE]{root mean squared error}
\end{acronym}


\begin{IEEEbiography}[{\includegraphics[width=1in,height=1.25in,clip,keepaspectratio]{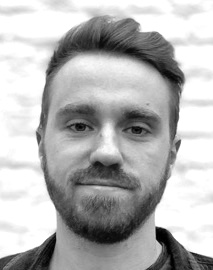}}]{Alexander Heinrich}
  received the B.Sc., M.Sc.\ degrees from the Technical University of Darmstadt, Darmstadt, Germany, in 2017 and 2019, respectively.

  He is a Ph.D. candidate with the Secure Mobile Networking Lab, Technical University of Darmstadt. His research is on the security and privacy of wireless protocols with a focus on location-aware systems. 

  Mr. Heinrich received several best demo awards and the best paper award at the ACM Conference on Security and Privacy in Wireless and Mobile Networks.  
\end{IEEEbiography}

\begin{IEEEbiography}[{\includegraphics[width=1in,height=1.25in,clip,keepaspectratio]{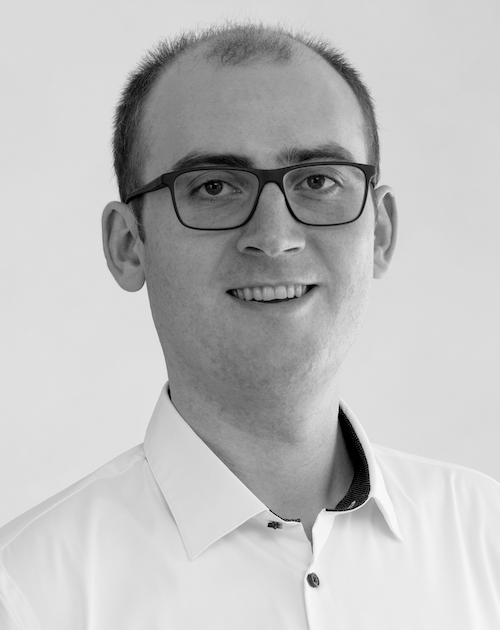}}]{Sören Krollmann}
  received the M.Sc.\ degree from  Technical University of Darmstadt, Darmstadt, Germany in 2022. 
  
  He worked as a scientific assistant with the Secure Mobile Networking Lab, Technical University of Darmstadt. His research focused on distance measurement using Ultra-Wide Band technology. 
  
  Mr. Krollmann currently works in industry as a cyber security consultant and supports customers on their way to more security.
\end{IEEEbiography}

\begin{IEEEbiographynophoto}{Florentin Putz}
  received the B.Sc.\ and M.Sc.\ degrees from the Technical University of Darmstadt, Darmstadt, Germany, in 2016 and 2019, respectively.

  He is a Ph.D. candidate with the Secure Mobile Networking Lab, TU Darmstadt. His research focuses on deployable and usable security mechanisms for smartphones and the Internet of Things.
  
  Mr.\ Putz received several awards, including the Best Student Award from TU Darmstadt's electrical engineering department as well as the KuVS Award 2020 for the best master's thesis in communications and distributed systems in Germany.
\end{IEEEbiographynophoto}

\begin{IEEEbiography}[{\includegraphics[width=1in,height=1.25in,clip,keepaspectratio]{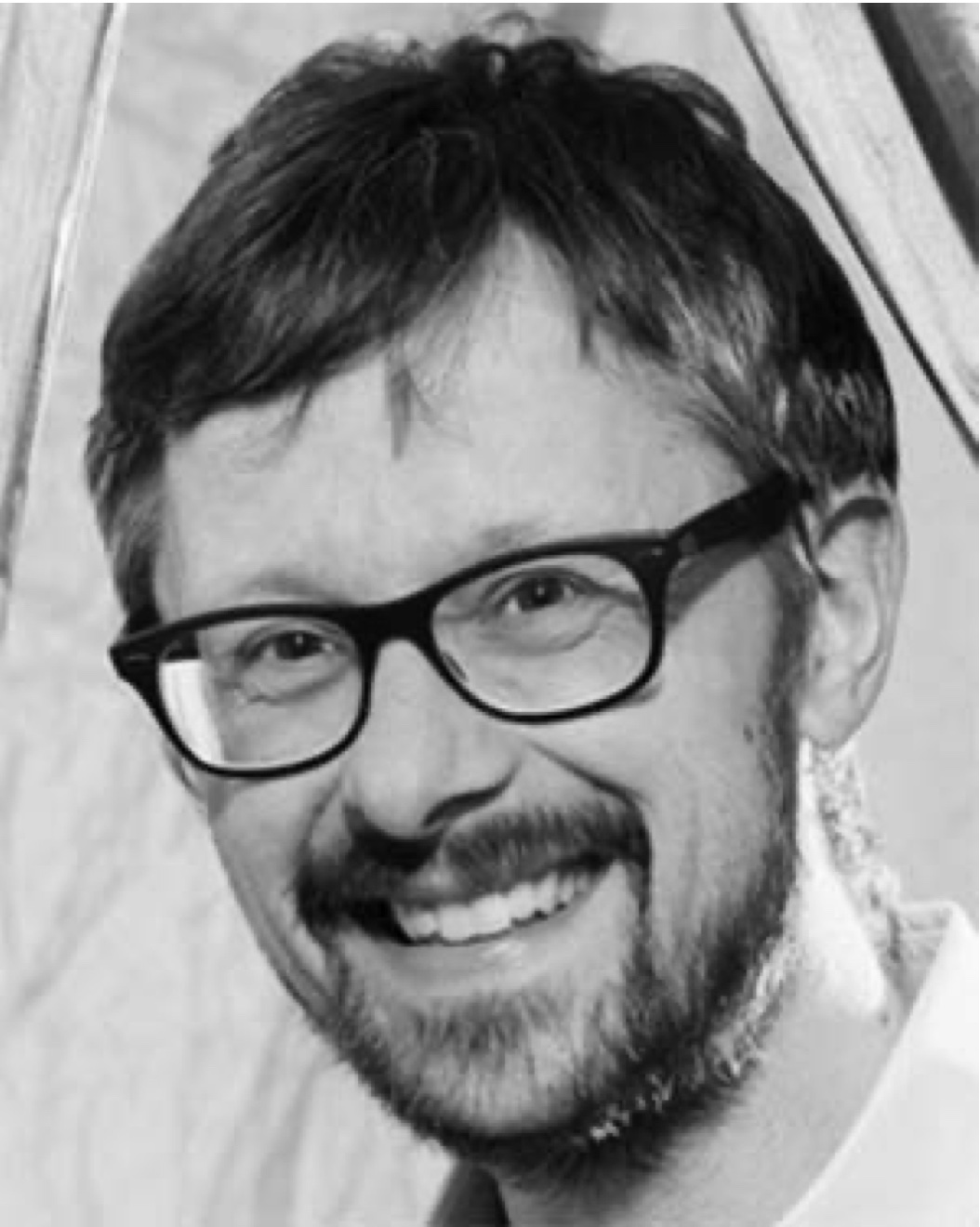}}]{Matthias Hollick}
  received his Ph.D.\ degree from TU Darmstadt, Darmstadt, Germany, in 2004.
  
  He is currently heading the Secure Mobile Networking Lab, Computer Science Department, Technical University of Darmstadt. He has been researching and teaching at TU Darmstadt, Universidad Carlos III de Madrid, Getafe, Spain, and the University of Illinois at Urbana–Champaign, Champaign, IL, USA. His research focus is on resilient, secure, privacy-preserving, and quality- of-service-aware communication for mobile and
  wireless systems and networks.  
\end{IEEEbiography}

\end{document}